\documentclass[conference]{IEEEtran}
\IEEEoverridecommandlockouts

\usepackage{cite}
\usepackage{amsmath,amssymb,amsfonts}
\usepackage{accents} 
\usepackage{graphicx}
\usepackage{textcomp}
\usepackage{xcolor}

\usepackage{bbm}
\usepackage{dsfont}
\usepackage{bbold}

\usepackage{tabularray}
\usepackage{stfloats}

\usepackage{float}

\usepackage{siunitx}
\usepackage{mathtools}
\usepackage{enumitem}

\usepackage{todonotes}

\usepackage{algorithm}
\usepackage{algpseudocode}

\allowdisplaybreaks 

\newcommand{\bfd}{\ensuremath{\mathbf{d}}}

\newcommand{\bfh}{\ensuremath{\mathbf{h}}}

\newcommand{\bfq}{\ensuremath{\mathbf{q}}}
\newcommand{\bfr}{\ensuremath{\mathbf{r}}}

\newcommand{\bfv}{\ensuremath{\mathbf{v}}}
\newcommand{\bfw}{\ensuremath{\mathbf{w}}}
\newcommand{\bfx}{\ensuremath{\mathbf{x}}}

\newcommand{\bfz}{\ensuremath{\mathbf{z}}}

\newcommand{\bfA}{\ensuremath{\mathbf{A}}}
\newcommand{\bfB}{\ensuremath{\mathbf{B}}}

\newcommand{\bfC}{\ensuremath{\mathbf{C}}}
\newcommand{\bfD}{\ensuremath{\mathbf{D}}}

\newcommand{\bfF}{\ensuremath{\mathbf{F}}}

\newcommand{\bfH}{\ensuremath{\mathbf{H}}}
\newcommand{\bfI}{\ensuremath{\mathbf{I}}}

\newcommand{\bfL}{\ensuremath{\mathbf{L}}}

\newcommand{\bfQ}{\ensuremath{\mathbf{Q}}}
\newcommand{\bfR}{\ensuremath{\mathbf{R}}}

\newcommand{\bfU}{\ensuremath{\mathbf{U}}}
\newcommand{\bfV}{\ensuremath{\mathbf{V}}}
\newcommand{\bfW}{\ensuremath{\mathbf{W}}}
\newcommand{\bfX}{\ensuremath{\mathbf{X}}}

\newcommand{\bfZ}{\ensuremath{\mathbf{Z}}}

\newcommand{\bbfz}{\ensuremath{\bar{\mathbf{z}}}}

\newcommand{\bbN}{\ensuremath{\mathbb{N}}}

\newcommand{\bbR}{\ensuremath{\mathbb{R}}}

\DeclareMathOperator*{\mean}{\mathbb{E}}				
\DeclareMathOperator*{\std}{std}				
\DeclareMathOperator*{\cov}{cov}			
\DeclareMathOperator*{\var}{var}			
\DeclareMathOperator*{\rank}{rank}			
\DeclareMathOperator*{\diag}{diag}			
\DeclareMathOperator*{\vecu}{vecu}				

\newcommand{\indfun}{\ensuremath{\mathds{1}}} 

\newcommand{\T}{\ensuremath{\text{T}}} 
\newcommand{\Ts}{\ensuremath{T_{\mathrm{s}}}}

\newcommand{\bfalpha}{\ensuremath{\boldsymbol{\alpha}}}

\newcommand{\bfmu}{\ensuremath{\boldsymbol{\mu}}}

\newcommand{\bfomega}{\ensuremath{\boldsymbol{\omega}}}
\newcommand{\bftau}{\ensuremath{\boldsymbol{\tau}}}
\newcommand{\bftheta}{\ensuremath{\boldsymbol{\theta}}}

\newcommand{\bfpi}{\ensuremath{\boldsymbol{\pi}}}
\newcommand{\bfPi}{\ensuremath{\boldsymbol{\Pi}}}
\newcommand{\bfnu}{\ensuremath{\boldsymbol{\nu}}}

\newcommand{\bfGamma}{\ensuremath{\boldsymbol{\Gamma}}}

\newcommand{\bfvarphi}{\ensuremath{\boldsymbol{\varphi}}}

\newcommand{\bfPhi}{\ensuremath{\boldsymbol{\Phi}}}
\newcommand{\bfepsilon}{\ensuremath{\boldsymbol{\epsilon}}}

\newcommand{\bfzeta}{\ensuremath{\boldsymbol{\zeta}}}

\RequirePackage{xspace}

\def\bfI{\mathbf I}
\def\bfV{\mathbf V}
\def\bfW{\mathbf W}
\def\bfX{\mathbf X}

\def\bfZ{\mathbf Z}

\def\real{\mathbb{R}}

\def\bfGamma{\mathbf \Gamma}

\def\calA{\mathcal{A}}
\def\calB{\mathcal{B}}
\def\calC{\mathcal{C}}
\def\calE{\mathcal{E}}
\def\calI{\mathcal{I}}
\def\calO{\mathcal{O}}
\def\calR{\mathcal{R}}

\def\bfnul{\mathbf{0}}

\def\T{^\intercal}
\def\sT{^{\!\intercal}}
\def\ssT{^{\!\!\intercal}}

\def\anni{\mathsf{am}}
\def\rank{\mathsf{rank}}

\def\tbfZ{\tilde{\bfZ}}

\def\rmU{{\mathrm{U}}}

\def\aO{n_{\mathsf{a}\raisebox{-0.45mm}{\scriptsize$\calO$}}}

\def\re{n_\calE}

\usepackage{accents}
\newcommand{\bbar}[1]{\accentset{\rule{.5em}{.15ex}}{#1}}

\newcommand{\bbfmu}{\bbar{\bfmu}}
\newcommand{\bbfx}{\bbar{\bfx}}
\newcommand{\bbfv}{\bbar{\bfv}}
\newcommand{\bbfw}{\bbar{\bfw}}
\newcommand{\bbfF}{\bbar{\bfF}}

\newcommand{\bbfQ}{\bbar{\bfQ}}
\newcommand{\bN}{\bbar{N}}

\newcommand{\te}{\mathit{t}}

\def\abs{\mathsf{abs}}

\def\diag{\mathsf{diag}}
\def\vertcat{\mathsf{vertcat}}
\def\vecu{\mathsf{vecu}}

\newcommand{\bfza}{\bfz_{\mathrm{a}}}
\newcommand{\bfthetaa}{\bftheta_{\mathrm{a}}}

\def\htheta{\hat{\theta}}

\newcommand{\doubletilde}[1]{\accentset{\approx}{#1}}

\begin{document}

\title{Noise Parameter Identification in Ensemble of Atomic Clocks
\thanks{This work was created with the support of the project "R\&D of Technologies for Advanced Digitization in the Pilsen Metropolitan Area (DigiTech)" No.: CZ.02.01.01/00/23\_021/0008436 co-financed by the European Union.}


}


\author{\IEEEauthorblockN{Jind\v{r}ich Dun\'{i}k, Oliver Kost, \\ Ladislav Kr\'{a}l, Ivo Pun\v{c}och\'{a}\v{r},}
\IEEEauthorblockA{\textit{University of West Bohemia} \\
Pilsen, Czech Republic\\
 \{dunikj,kost,ladkral,ivop\}@fav.zcu.cz}
\and
\IEEEauthorblockN{Ond\v{r}ej Daniel}
\IEEEauthorblockA{\textit{HULD s.r.o.} \\
Prague, Czech Republic \\
ondrej.daniel@huld.io}
\and
\IEEEauthorblockN{Simona Circiu, Bernardino Quaranta}
\IEEEauthorblockA{\textit{ESA/ESTEC, Noordwijk} \\
Noordwijk, The Netherlands \\
\{bernardino.quaranta,simona.circiu\}@esa.int}
}


\maketitle

\begin{abstract}

The paper introduces a parameter identification framework for atomic clock ensembles, where individual clocks are represented by second-order linear stochastic state-space model. The proposed procedure relies exclusively on differential measurements, namely the pairwise phase differences between a designated pivot clock and the remaining ensemble members. The identification targets key parameters for each clock: white and random-walk frequency noise variances, constant frequency drift, and (co)variance of the measurement noise. To identify these parameters, two distinct identification methods are proposed. Their estimation accuracy is rigorously demonstrated via numerical simulations and validated using empirical data from atomic hydrogen maser (AHM) clocks at ESTEC Timing Laboratory.
\end{abstract}



\section{Introduction}
Precise timekeeping acts as the invisible backbone of modern infrastructure, synchronizing everything from global satellite navigation to high-frequency trading and distributed telecommunications \cite{Allan1997,Galleani2010:IEEECSM}. To guarantee extreme stability and robustness against individual hardware failures, high-quality time scales rely on ensembles of atomic clocks. While classical aggregation methods successfully reduce overall frequency instability, the next frontier in precision timekeeping demands a shift toward dynamic, model-based estimation approaches.

Parametric models such as state-space models can significantly enhance the stability and integrity of a time scale by incorporating prior knowledge of clock behavior. However, the efficacy of these model-based algorithms is fundamentally bounded by the accuracy of their underlying stochastic models. Real-world atomic clocks exhibit complex, often correlated noise profiles that are notoriously difficult to isolate and quantify. Therefore, the robust identification of noise parameters within an ensemble is a critical prerequisite. Overcoming this parameter identification challenge enables optimal weighting strategies and the timely detection of anomalies, bridging the gap between theoretical clock models and maximum real-world performance.

Although stochastic clock modeling and noise identification have been studied for decades, reliable estimation remains difficult in practical scenarios characterized by limited data records, correlated observations, and incomplete observability of individual clocks. Moreover, several theoretical aspects of the problem remain open, particularly regarding identifiability, sub-space unobservability, estimator consistency, or the statistical characterization of estimation uncertainty under realistic operating conditions. Furthermore, existing approaches predominantly address isolated aspects of the problem, seldom offering a unified framework that is simultaneously comprehensive, computationally efficient, and practically applicable.

\section{Related Works}
An ensemble of atomic clock is an underlying component of many applications that range from timekeeping to positioning. A successful operation of an ensemble relies on several algorithms that provide anomaly monitoring, clock combination, or clock steering. Pioneering clock combination algorithms are adaptive and only little information about clocks in the ensemble is required to initialize these algorithms~\cite{Weiss1991:Metrologia}. On the other hand, there are  algorithms which assume that a complete model of the clock ensemble is available~\cite{Stein1988:cp,Galleani2010:IEEECSM}. Therefore, there has been a concentrated effort devoted to clock ensemble identification. Since the the underlying dynamic is given by law of physics, the main effort is concentrated on estimating the parameters of the noises.

The parameter estimation method is tightly related to a selected model. Although there are some works focusing on ARIMA models~\cite{Barnes1988,Audoin1993}, most of the works related to the clock ensemble identification are based on the state space models \cite{ZuTa05:IEEE} that are physically interpretable. A particular attention was devoted to the estimation of the frequency drift~\cite{Barnes1983b} from time deviations, the first order difference of time deviations (i.e. average normalized frequency deviations), and the second order difference of time deviations.

One approach to noise parameter identification is based on the maximum likelihood estimation method~\cite{Tryon1983,Jones1983,Barnes1983a}, where estimates of unknown model parameters are found to maximize the likelihood of measurements~\cite{Tsang1981:IEEETAC}. The likelihood of measurement is written using a chain rule as the product of measurement predictive probability density functions that are obtained using the Kalman filter. Missing measurements can be handled rather easily and apparent measurement outliers can by removed while constructing the likelihood. A potential drawback of the maximum likelihood estimation is the necessity of relying on an iterative optimization method as the likelihood is a nonlinear function of model parameters. To avoid negative variance estimates, either explicit constrained optimization is performed or variances are replaced by positive functions of auxiliary variables.

Another approach to noise parameter identification is based on the three-cornered-hat method that allows variances of three independent random variables to be estimated even if only samples of their pair-wise differences are available. This idea was applied to Allan variance (AVAR) plots under the assumption of noise-free measurements and independent internal noises of clocks in~\cite{Gray1974}. If more than three clocks are available, triads are formed and parameter estimated. Later on, the three-cornered-hat method has been extended to m-cornered-hat method~\cite{Greenhall:87} where a comparison with maximum likelihood method is given for a rather simplified ensemble model. The m-cornered-hat method also suffers from possibility of getting negative variance estimates. Several reasons for this issue are discussed in literature. The variance estimate uncertainty due to insufficient number of measurements is suspected in~\cite{Gray1974}. In~\cite{Vernotte2016}, the measurement noise that influences the AVAR plot for short averaging intervals is suspected as a possible reason and the Groslambert covariance (Allan covariance (ACOV))~\cite{Fest1983:IEEETIM} is used instead of the AVAR under the assumption that the measurement noises are uncorrelated to make covariance estimates independent of measurement noise. A more in-depth discussion of possible reasons was given in~\cite{Premoli1993:IEEETIM}. It was argued that most commonly considered reasons in literature are not plausible and conclude that the clock noises should be admitted to be correlated. However, it makes the variance estimation problem ambiguous. These issue is discussed  for the three-cornered-hat method in~\cite{Premoli1993:IEEETIM} and this discussion is extended to the m-cornered-hat method in~\cite{Tavella1993:Metrologia} and~\cite{Torsaco2000:IEEETUFFC}.

\section{Contribution} \label{sec:Contribution}
This paper addresses the problem of identifying noise parameters in a clock ensemble following the standard approach of utilizing differential measurements between a selected pivot clock and the remaining clocks. The main objective is to develop an effective identification method tailored to this measurement configuration, enabling reliable estimation of ensemble noise characteristics under realistic observational constraints. 

Two noise parameter estimation methods are presented in this paper. The first one represents the extension of an underlying correlation approach for noise identification presented in~\cite{Mehra1970:IEEETAC}. Note that this approach was considered in the context of the clock ensemble identification in~\cite{Stein1989} but with unrealistic assumptions regarding the direct measurement of time deviations. The second presented methods builds upon the Hadamard covariance (HCOV) and m-cornered-hat method. This is similar to the concept in \cite{Vernotte2016,dunikAll:26}, where, however, the authors use Allan covariance and highlight its advantages over the conventional three-cornered-hat method.
Both designed identification methods represent an original approach; they are broadly applicable to a wide class of ensemble atomic clocks. Moreover, their design is transparent and builds on the strong theoretical foundation of linear estimation.

\section{A clock and ensemble model definition} \label{sec:ensembleDef}
\IEEEpubidadjcol 
The paper follows the standard modeling on atomic clock \cite{ZuTa05:IEEE}. It is assumed that the $i$-th clock of an ensemble can be described at a time step $k = 0,1,2\ldots,N$ by a second-order discrete-time state-space model\footnote{Notation: A symbol $x^{(i)}$ denotes a parameter or a variable belonging to the $i$-th clock, $x_{i,k}$ is the $i$-th component of vector $\bfx$ in the time instant $k$, $x_{i}$ is the $i$-th element of the vector $\bfx$, $X_{ij}$ is the $(i,j)$-th element of the matrix $\bfX$, $\bfI_n \in \bbR^{n \times n}$ is an $n$-dimensional identity matrix, $\bfx_{i:j} = [x_{i}, x_{i+1}, \ldots, x_{j} ]^{\T}$ is a sequence of $x$ from step $i$ to step $j$, $\indfun_{m}\in \bbR^m$ is a vector of ones, $\mathbf{0}_{m\times n} \in \bbR^{m \times n}$ is the zero matrix, $\hat{x}$ is an estimate of $x$, $\otimes$ is the Kronecker product, $\vertcat$ is a vertical concatenating operation, $\diag$ is a block diagonal concatenation operator, $\mean$ is the expectation operator, and $\var$, $\cov$ are operators of variance and covariance. The symbols $\bbR^{+}$ and $\bbR^{++}$ denote real non-negative and real positive numbers, respectively.}
\begin{align}
    \bfx_{k+1}^{(i)} & = \bfF_1 \bfx_k^{(i)} + \bfmu^{(i)} + \bfw_k^{(i)},  \label{eq:clk1}\\
    \Delta_k^{(i)} & = \bfh_{1}^{\T} \bfx_k^{(i)},  \label{eq:clk2}
\end{align}
where $\bfx_k^{(i)} = [x_{1,k}^{(i)}, \ x_{2,k}^{(i)}]^{\T} \in \bbR^2$ is a state consisting of a time deviation $x_{1,k}^{(i)}\in\bbR$ and a random walk process $x_{2,k}^{(i)}\in \bbR$,  $\bfmu^{(i)}\in\bbR^{2}$ is a drift term,  $\bfw_k^{(i)}\in\bbR^{2}$ is a zero mean white Gaussian state noise with covariance matrix (CM) $\bfQ^{(i)}\in\bbR^{2 \times 2}$, and $\Delta_{k}^{(i)}\in\bbR$ is a time deviation. The matrices $\bfF_{1}$, $\bfQ^{(i)}$ and vectors $\bfh_{1}$, $\bfmu^{(i)}$ are given as
\begin{equation}
    \begin{aligned}
        \bfF_1 =
    \begin{bmatrix} 
        1 & \Ts \\
        0 & 1 
    \end{bmatrix}&,\ 
    \bfh_{1}^{\T} = 
    \begin{bmatrix} 
        1 & 0 
    \end{bmatrix},\ 
    \bfmu^{(i)} = d^{(i)} 
    \begin{bmatrix} 
        \Ts^2/2 \\
        \Ts 
    \end{bmatrix},\\
    \bfQ^{(i)}& = 
    \begin{bmatrix} 
        q_1^{(i)} \Ts + q_2^{(i)} \Ts^3/3 & q_2^{(i)} \Ts^2/2 \\ 
            q_2^{(i)} \Ts^2/2 & q_2^{(i)} \Ts
    \end{bmatrix}, 
    \end{aligned}
\end{equation}
where $\Ts\in \bbR^{++}$ is a sampling period in seconds, $d^{(i)} \in \bbR$ is a frequency drift, and $q_1^{(i)} \in \bbR^{++}$ and  $q_2^{(i)}\in \bbR^{++}$ denote intensities of original continuous-time stochastic processes that represent the white noise frequency modulation and random walk frequency modulation, respectively.

The model of ensemble of $n$ clocks can be defined as
\begin{align}
    \bfx_{k+1} &= \bfF \bfx_k + \bfmu + \bfw_k, \label{eq:ensemble_state} \\
    \bfz_k &=  \bfH \bfx_k + \bfv_k,\label{eq:ensemble_meas}
\end{align}
where $\bfx_{k} \in \bbR^{2n}$ is an ensemble state, $\bfmu\in\bbR^{2n}$ is ensemble drift,  $\bfw_{k} \in \bbR^{2n}$ is an ensemble state noise with zero mean and CM $\bfQ\in\bbR^{2n \times 2n}$, $\bfz_k \in \bbR^{n_z}$ is a measurement containing $n_z = n-1$ differences of time deviations of a pivot clock and all other clocks, and $\bfv_{k} \in \bbR^{n_z}$ is a white Gaussian measurement noise with zero mean and CM $\bfR \in \bbR^{n_z \times n_z}$. The vectors and matrices are given as
\begin{align}
    \bfx_k &= \vertcat\!\left( \bfx_k^{(1)}, \bfx_k^{(2)}, \dots, \bfx_k^{(n)} \right),\\
    \bfw_k &= \vertcat\!\left( \bfw_k^{(1)}, \bfw_k^{(2)}, \dots, \bfw_k^{(n)} \right),\\
    \bfz_k & = \left[ z_{1,k}, z_{2,k},\ldots, z_{n_z,k} \right]^{\T}, \\
    \bfF &= \bfI_{n} \otimes \bfF_{1}, \quad \bfH = \bfC \left( \bfI_{n} \otimes \bfh_{1}^{\T} \right),\\
    \bfmu & = \vertcat\!\left( \bfmu^{(1)}, \bfmu^{(2)}, \dots, \bfmu^{(n)} \right),\\
    \bfQ &= \diag\!\left( \bfQ^{(1)}, \bfQ^{(2)}, \dots, \bfQ^{(n)} \right),
\end{align}  
where $\bfC = \left[ -\mathds{1}_{n_z}, \bfI_{n_z} \right]\in\bbR^{n_z\times n}$ is a matrix that specifies the first clock in the ensemble to be the pivot clock.

\section{Problem Formulation} \label{sec:ProblemFormualtion}
In principle, the problem does not permit an unambiguous estimation of the frequency drift of all clocks. Therefore, a common approach is to assume zero drift for the pivot clock and estimate the remaining drifts relative to it, i.e., the estimate the drift differences between the pivot and the other clocks in the ensemble.
Thus, let's define a vector of unknown model parameters as
\begin{equation}
     \bftheta = \left[ \bfq_1^{\T}, \bfq_2^{\T}, \bfr^{\T}, \bfd^{\T} \right]^{\sT}  = \left[\bftheta_a^{\T}, \bfd^{\T}  \right]^{\sT},
\end{equation}
where $\bftheta \in \bbR^{n_\theta}$ with $n_\theta \!=\! (n^2+5n-2)/2$ denoting the number of the  unknown parameters, $\bfq_1^{\T} \!=\!  \left[ q_1^{(1)}, q_1^{(2)}, \dots, q_1^{(n)}\right]\in \bbR^{n}$, $\bfq_2^{\T} \!=\!  \left[ q_2^{(1)}, q_2^{(2)}, \dots, q_2^{(n)}\right]\!\in\! \bbR^{n}$,  $\bfr^{\T} \!=\! \vecu(\bfR) \!\in\! \bbR^{n_r}$ collects elements of the upper triangular CM $\bfR$ with $n_r = (n^2-n)/2$ and $\bfd^{\T}  \!=\! \left[  d^{(2)}, d^{(3)} \dots, d^{(n)} \right]\in \bbR^{n-1}$. The vector $\bftheta_a \in \bbR^{n_{\theta_a}}$ with $n_{\theta_a} = n(n+3)/2$ collects parameters $q_1^{(i)}$, $q_2^{(i)}$ and $R_{ij}$. 

\textbf{Goal:} Given a model ($\bfF$, $\bfH$, $\Ts$)  and an available sequence of measurements  $\bfz_{0:N}$, the unknown model parameters $\bftheta$ are estimated as
\begin{align} \label{eq:optimProblem}
    \hat{\bftheta} = \arg \min_{\bftheta} J(\bftheta, \bfz_{0:N}), 
\end{align}
where $J\!:\!\bbR^{n_{\theta}\times (N+1)n_{z}} \!\mapsto\! \bbR^{++}$ is a designer-selected criterion. 

\section{Identification Methods} \label{sec:IdentificationMethods}
In this section two methods for estimating the model parameters are presented. 
Contrary to the  standard maximum likelihood methods, both proposed methods aim to avoid solving large non-linear optimization problems by using a suitable transformation of available measurements.

\subsection{Method based on Hadamard covariance (HCOV)} \label{subsec:AVAR_Methods}
The basic idea is to use HCOV, whose unique properties enable (i) transformation into a linear optimization problem and (ii) elimination of drift effects, allowing the drift estimation problem to be treated separately.

\subsubsection{HCOV Transformation}
The HCOV can be defined for measurement sequences $\{z_{i,k}\}_{k=0}^{\infty}$ and $\{z_{j,k}\}_{k=0}^{\infty}$ as
\begin{multline} \label{eq:acov_meas}
   \sigma^2_{i,j} (\tau_m)  = \frac{1}{6\tau_m^2} \mean \left\{ (z_{i,k+3m} - 3z_{i,k+2m} + 3z_{i,k+m} - z_{i,k}) \right.\\ \times \left. (z_{j,k+3m} - 3z_{j,k+2m} + 3z_{j,k+m} - z_{j,k}) \right\},
\end{multline} 
where $\tau_m=m \Ts \in \bbR^{++}$ is an averaging time with $m \in \bbN$. The ensemble model~\eqref{eq:ensemble_state}--\eqref{eq:ensemble_meas} and the theory of stochastic processes can be employed (see Appendix \ref{sec:appendixA}) to get an analytical expression that relates the model parameters $\bftheta$ to the HCOV $\sigma^2_{i,j} (\tau_m)$. If $i \neq j$, the HCOV is given as
\begin{equation} \label{eq:acovAnalytic1}
    \sigma^2_{i,j} (\tau_m) =  \frac{q_1^{(1)}}{\tau_m} + \frac{q_2^{(1)}}{6}\tau_m +   \frac{10R_{ij}}{3\tau_m^2}.
\end{equation}
If $j = i$ the HCOV reduces to Hadamard variance (HVAR) and it is given as
\begin{align} \label{eq:acovAnalytic2}
    \sigma^2_{i,i} (\tau_m) = \frac{q_1^{(1)} \!+\! q_1^{(i+1)}}{\tau_m} \!+\! \frac{q_2^{(1)} \!+\! q_2^{(i+1)}}{6}\tau_m  \!+\!   \frac{10 R_{ii}}{3\tau_m^2}.
\end{align}

The expressions \eqref{eq:acovAnalytic1}--\eqref{eq:acovAnalytic2} show that for a fixed $\tau_m$, the HCOV/HVAR are linear functions of all model parameters except for the drifts on which they are independent. 

\subsubsection{Regression model}
Assuming that HCOV estimates $\hat{\sigma}^2_{i,j}\in\bbR^{\te}$ are computed using \eqref{eq:acov_meas} where the expectation operator is evaluated as the sample mean with measurements $\bfz_{0:N}$ for $\te$ different averaging times
$\bar{\bftau} =  \left[ \bar{\tau}_{1}, \bar{\tau}_{2}, \dots, \bar{\tau}_{\te}\right] \in \bbR^{\te}$, where
$\bar{\tau}_{p} = m_{p}\Ts$ with $m_{p}\in\{1,2,\ldots, N \}$. 
Then, it is possible to use \eqref{eq:acovAnalytic1}--\eqref{eq:acovAnalytic2} to write the following linear regression model
\begin{equation} \label{eq:compact}
    \bfza = \bfPhi \bfthetaa + \bfepsilon,
\end{equation}
where $\bfza\in \bbR^{n_{z_a}}$ is a vector given as
\begin{align} \label{eq:zaDef}
    \!\!\!\!\bfza =
   \vertcat \left( \hat{\sigma}^2_{1,1}, \hat{\sigma}^2_{2,2}, \ldots, \hat{\sigma}^2_{n_z,n_z},
   \hat{\sigma}^2_{1,3},\ldots, \hat{\sigma}^2_{n_z-1,n_z} \right)\!\!\!\! 
\end{align}
with $n_{z_a} = n (n-1)\te/2$.  The regression matrix $\bfPhi \in \bbR^{n_{z_a} \times n_{\theta_a}}$ is defined as
\setlength{\arraycolsep}{1.5pt}
\begin{align}
\bfPhi \!=\!
\begin{bmatrix}
    \bfA_1
    \! \otimes \! \left( \bfvarphi
    \left[
     \begin{smallmatrix}
          1  \\  0 \\ 0
     \end{smallmatrix} 
    \right]  \right)\!, 
&
    \bfA_1
    \! \otimes \! \left( \bfvarphi
    \left[
     \begin{smallmatrix}
          0 \\  1 \\ 0
     \end{smallmatrix} 
    \right]  \right)\!,
&
    \bfA_2
    \! \otimes \! \left( \bfvarphi
    \left[
     \begin{smallmatrix}
          0 \\  0 \\ 1
     \end{smallmatrix} 
    \right]  \right) 
    \end{bmatrix}\!,
\end{align}
where
\begin{gather}
    \bfvarphi = \vertcat ( \varphi_1, \varphi_2, \ldots, \varphi_\te), \quad 
    \varphi_i = \left[ 
    \begin{smallmatrix} \frac{1}{\bar{\tau}_i}, & \frac{\bar{\tau}_i}{6}, &\frac{10}{3\bar{\tau}_i^2} \end{smallmatrix}
    \right],\\
\bfA_1 =  \label{eq:A1A2def}
\left[
     \begin{smallmatrix}
          \indfun_{n_z} & \bfI_{n_z} \\ 
          \indfun_{n_f} & \mathbf{0}_{n_f \times n_z} 
     \end{smallmatrix} 
\right], 
\quad
\bfA_2 = 
\left[
     \begin{smallmatrix}
          \bfI_{n_z} & \mathbf{0}_{n_z \times n_f}  \\ 
          \mathbf{0}_{n_f \times n_z} & \bfI_{n_f}
     \end{smallmatrix} 
\right],
\end{gather} 
and $n_f = n_z(n_z-1)/2$.
The HCOV estimation error $\bfepsilon \in \bbR^{n_{z_a}}$ has zero mean. Since the true CM of $\bfepsilon$ is difficult to obtain, it is approximated by a diagonal CM $\bfR_{\theta_a} \in \bbR^{n_{z_a} \times n_{z_a}}$
where each diagonal element represents the corresponding variance. 
Specifically, the variance of HCOV estimate $\hat{\sigma}^2_{i,j} (\bar{\tau}_{p})$ can be obtained \cite{muirhead:82} from the properties of the Wishart distribution or as the variance of the sample covariance, where it is assumed that the signals are jointly Gaussian. Then, it can be written
\begin{equation} \label{eq:varOfAvar2}
    \var \left\{\hat{\sigma}^2_{i,j} (\bar{\tau}_p) \right\} \approx \left[ \left( \hat{\sigma}^2_{i,j}  (\bar{\tau}_{p}) \right)^2 +  \hat{\sigma}^2_{i,i}  (\bar{\tau}_{p})  \hat{\sigma}^2_{j,j}  (\bar{\tau}_{p}) \right]  / \nu. 
\end{equation} 
This process involves two key approximations: substituting true (co)variances by their estimated values and reducing the degrees of freedom $\nu$ to account for correlations in the measurement sequence.  As $\nu$ depends on the underlying noise process \cite{Riley:94}, adopting a random-walk frequency modulation (RWFM) model serves as a conservative choice with 
\begin{align} \label{eq:nu_def}
\nu \approx N^2/\left(31/30 \!\cdot\! Nm - 17/28 \!\cdot\! m^2 \right).
\end{align}

The estimate of the unknown parameters $\bfthetaa$ is obtained by solving the linear minimization problem \eqref{eq:optimProblem} with the criterion
\begin{equation} \label{eq:J2_def}
  J(\bftheta_a, \bfz_{0:N}) = ||\bfR^{-1/2}_{\theta_a} (\bfza - \bfPhi \bftheta_a) ||^2. 
\end{equation}
Note that the number of averaging times $\te$ and their particular values represent design parameters.


\subsubsection{Drift Estimation}
Now let us return to the problem of drift estimation. The several approaches can be discussed:
\begin{itemize}[topsep=0pt, itemsep=0pt, leftmargin=*]
    \item A regression analysis on the phase time-deviation sequence or its first and second differences using quadratic, linear, or constant models, respectively \cite{Riley:94}. 
    \item Modification of HCOV method, where HCOV is replaced by Allan covariance \cite{dunikAll:26}. This approach leads to a nonlinear optimization problem and, moreover, does not enable determination of the drift sign.  
    \item Kalman filter based estimation, where the state-space model \eqref{eq:clk1}--\eqref{eq:ensemble_meas} can be re-casted such that the second order clock model \eqref{eq:clk1}-\eqref{eq:clk2} is replaced by the third order model, where the third state is the unknown constant drift. 
    \item  Measurement difference method (MDM) approach (see below Sec. \ref{fDriftSec}).
\end{itemize}
In the considered scenario of long and correlated data records, estimation of the drift $d^{(i+1)}$ is performed by the first option. Concretely, a regression analysis is applied to the second difference of the $i$-th component of $\bfz_k$
\begin{equation}\label{eq:2ndDiff}
\Delta^2 z_{i,k} = (z_{i,k+2} - 2z_{i,k+1} + z_{i,k})/\Ts.
\end{equation}
Then, for given the state-space model \eqref{eq:clk1}--\eqref{eq:clk2}, the appropriate regression model is 
\begin{equation}
\Delta^2 z_{i,k} = d^{(i+1)} + \varepsilon_{i,k}.
\end{equation}
Using the analogical procedure from Appendix~\ref{sec:appendixA}, it can be easily shown that $\{ \varepsilon_{i,k} \}$ is a moving average process of order~2 and satisfies 
\begin{align}
\!\!\!\!\cov(& \varepsilon_{i,k}) \!=\! \frac{2(q_1^{(i)} \!\!+\! q_1^{(i+1)})}{\Ts} \! + \!\frac{2(q_2^{(i)}\!\!+\!q_2^{(i+1)})}{3}\Ts\! +\! \frac{6R_{ii}}{\Ts^2}, \label{eq:covd1}\\
\!\!\!\!\cov(&  \varepsilon_{i,k},   \varepsilon_{i,k+1}) \!=\! -\frac{q_1^{(i)} \!\!+\! q_1^{(i+1)}}{\Ts} \! + \!\frac{q_2^{(i)}\!\!+\!q_2^{(i+1)}}{6}\Ts\! -\! \frac{4R_{ii}}{\Ts^2},\! \label{eq:covd2}\\
\!\!\!\!\cov(&  \varepsilon_{i,k},   \varepsilon_{i,k+2}) \!= \! R_{ii}/\Ts^2. \label{eq:covd3} 
\end{align}
It follows that the original strongly correlated integration process $\bfz_k^{(i)}$ was transformed into a process with a very short correlation length, which significantly simplifies the following linear optimization problem
\begin{equation} \label{eq:optimProblemDrift}
    \hat{d}^{(i+1)}= \arg \min_{d}  ||(\bfR_d^{(i)})^{-1/2} (\Delta^2 \bfz_{k}^{(i)} - d) ||^2, 
\end{equation}
where $\bfR_d^{(i)} \in \bbR^{(N-2) \times (N-2)}$ is an error CM. It has a band sparse structure with the $(I,J)$ term given as 
\begin{multline}\label{eq:wIJdef}
    R^{(i)}_{d_{I,J}} = \cov( \varepsilon_{i,k})\delta_{I,J} + \cov( \varepsilon_{i,k},  \varepsilon_{i,k+1}) \delta_{I,J\pm1} \\ + \cov( \varepsilon_{i,k}, \varepsilon_{i,k+2}) \delta_{I,J\pm2}, 
\end{multline}
where $\delta_{i,j}$ is a Kronecker delta operator. Note that \eqref{eq:covd1}--\eqref{eq:covd3} can be realized by replacing the true noise parameters with their estimates. As a result, the drift estimates depend on $\hat{\bftheta}_a$.


Finally, the HCOV method is summarized in Algorithm~\ref{alg:HCOV}.

\begin{table}[ht!] 
    \vspace{-5mm}
    \centering
        \begin{algorithm}[H] 
        \caption{The estimation algorithm based on HCOV} \label{alg:HCOV}
            \begin{algorithmic}[1]
                \Require $\bfz_{1:N}$, $\bar{\bftau}$, 
                \For{$i=1,\ldots,n_z$ and $j=i,\ldots,n_z$}
                    \State Estimate $\hat{\sigma}^2_{i,j} (\bar{\bftau})$ as in \eqref{eq:acov_meas} 
                    \State Compute $\var(\hat{\sigma}^2_{i,i} (\bar{\bftau}))$  and  $\var(\hat{\sigma}^2_{i, j} (\bar{\bftau}))$ as in \eqref{eq:varOfAvar2}
                    \State Store the results from Steps 2--3
                \EndFor
                \State Construct $\bfz_a$, $\bfR_{\theta_a}$ and $\bfPhi$ as in \eqref{eq:compact}--\eqref{eq:nu_def}.
                \State Estimate $q_1^{(i)}$, $q_2^{(i)}$, $R_{ij}$ as in \eqref{eq:optimProblem}, \eqref{eq:J2_def}.
                \For{$i=1,\ldots,n_z$}
                 \State Construct $\Delta^2 \bfz_{k}^{(i)}$, $\bfR_d$ as in \eqref{eq:2ndDiff}, \eqref{eq:covd1}--\eqref{eq:covd3} and \eqref{eq:wIJdef} 
                   \State Estimate $d^{(i+1)}$ using the procedure \eqref{eq:optimProblemDrift}
                \EndFor
                \Ensure Estimated parameters $\hat{\bftheta}$ and their CMs.
            \end{algorithmic}
        \end{algorithm}
     \vspace{-6mm}
\end{table}
\subsection{Measurement Difference Method}\label{subsec:MDM_Methods}
The unique properties of MDM stem from the definition of the residue, which can be calculated from available measurements and shown to be a linear function of the immeasurable state and measurement noises \cite{KoDuPuSt:26}.

\subsubsection{The Sampling Period and Resampled Model}\label{sec:resMDM}
Based on an initial analysis of the measurements and insights from the ACOV/HCOV techniques, it can be shown that the sampling period affects the extent to which individual parameters are reflected in the measurements and, consequently, the quality of the estimates of these parameters. This phenomenon is presented in Section \ref{influenceTs}.
Therefore, let's define the resampled model of ensemble of $n$ clocks from Section~\ref{sec:ensembleDef} as
\begin{subequations}\label{eq:ensemble_xx}
\begin{align}
    \!\!\!\bbfx_{k+1} \!&=\! \overbrace{\bfI_{n} \!\otimes\!\! \begin{bmatrix} 
        1 & \tau_m \\
        0 & 1 
    \end{bmatrix}}^{\bbfF} \bbfx_k + 
    \overbrace{
    \begin{bmatrix} \bfnul_{1 \times (n-1)}\\ \bfI_{n-1}\end{bmatrix}\!\!\otimes\!\!\begin{bmatrix}\tau_m^2/2\\ \tau_m
    \end{bmatrix}\!
    \bfd
    }^{\bbfmu} + \bbfw_k, \label{eq:ensemble_statexx} \!\!\!
    \\
    \bbfz_k &=  \bfH \bbfx_k + \bbfv_k,\label{eq:ensemble_measxx}
\end{align}
\end{subequations}
for $k = 0,1,\ldots,(\bN\!-\!1)$, where $\bN\!=\!(\lfloor N/m\rfloor\!+\!1)$ is a total number of downsampled measurements, $\tau_m \!=\! m\Ts \in \bbR^{++}$ is a resampling period\footnote{Note that in Section \ref{subsec:AVAR_Methods}, the same variable $\tau_m$ was named \textit{averaging time}. However, within the HCOV identification method, it had a different meaning and application compared to the MDM method.} 
in seconds, $m \in \bbN$ determines the multiple of the original sampling period $\Ts$, and the relations also hold:
\begin{subequations}
\begin{align}
    \bbfz_k&=\bfz_{m k} , \quad \bbfv_k=\bfv_{mk} , \quad  \cov(\bbfv_k) = \bfR,
    \\
    \bbfQ &= \cov(\bbfw_k) = \diag \left( \bbfQ^{(1)}, \bbfQ^{(2)}, \dots, \bbfQ^{(n)} \right),
    \\
    \bbfQ^{(i)}& = 
    \begin{bmatrix} 
        q_1^{(i)} \tau_m + q_2^{(i)} \tau_m^3/3 & q_2^{(i)} \tau_m^2/2 \\ 
            q_2^{(i)} \tau_m^2/2 & q_2^{(i)} \tau_m
    \end{bmatrix}.
\end{align}
\end{subequations}

\subsubsection{Augmented Measurement Vector}
Let's define the augmented measurement vector as \cite{KoDuPuSt:26}
\begin{align}
	\bfZ_k = \calO\bfx_k + \bfGamma\bfW_k + \bfV_k + \bfGamma\bfU \label{Z}
\end{align}
for $k=0,\ldots,(\bN\!-\!L)$, where $L\!\geq\!1$ is a user-defined parameter, $\bfZ_k\in\real^{Ln_z}$, $\bfW_k\in\real^{(L-1)n_x}$, $\bfV_k\in\real^{Ln_z}$ $\calO\in\real^{Ln_z\times n_x}$, $\bfGamma\in\real^{Ln_z\times (L-1)n_x}$, $\bfU\in\real^{(L-1)n_x}$, and $\Upsilon_{\bfd}\in\real^{(L-1)n_x \times (n-1)}$ are defined 
as 
\begin{subequations}\label{eq:matrices}
\begin{align}
    \!\!\!\bfZ_k\!&\triangleq\!\!\left[\begin{smallmatrix} \bbfz_k \\ \bbfz_{k+1} \\ \bbfz_{k+2} \\ \vdots \\ \bbfz_{k+L-1} \end{smallmatrix}\right]\!\!,\ 
    \bfW_{k}\!\triangleq\!\!\left[\begin{smallmatrix} \bbfw_k \\ \bbfw_{k+1} \\ \bbfw_{k+2} \\ \vdots \\ \bbfw_{k+L-2} \end{smallmatrix}\right]\!\!,\ 
    \bfV_k\!\triangleq\!\!\left[\begin{smallmatrix} \bbfv_k \\ \bbfv_{k+1} \\ \bbfv_{k+2} \\ \vdots \\ \bbfv_{k+L-1} \end{smallmatrix}\right]\!\!,\!\!
    \\
    \!\!\calO\!&\triangleq\!\!\left[\begin{smallmatrix} \bfH \\ \bfH\bbfF \\ \bfH\bbfF^2 \\ \vdots \\ \bfH\bbfF^{L-1} \end{smallmatrix}\right]\!\!,\ 
    \bfGamma\!\triangleq\!\!\left[\begin{smallmatrix}
        \bfnul_{n_z\times n_x} & \bfnul_{n_z\times n_x}& \cdots & \bfnul_{n_z\times n_x} 
		\\ \bfH & \bfnul_{n_z\times n_x} & \cdots & \bfnul_{n_z\times n_x} 
		\\ \bfH\bbfF & \bfH & \cdots & \bfnul_{n_z\times n_x}
		\\ \vdots & \vdots & \ddots & \vdots
		\\ \bfH\bbfF^{L-2} & \bfH\bbfF^{L-3} & \cdots& \bfH
    \end{smallmatrix}\right]\!\!,
    \\
    \bfU\!&\triangleq\!\left[\begin{smallmatrix} \bbfmu \\ \bbfmu \\ \bbfmu \\ \vdots \\ \bbfmu \end{smallmatrix}\right]
    \!=
    \underbrace{\!\left(\indfun_{L-1}\otimes\!\begin{bmatrix} \bfnul_{1 \times (n-1)}\\ \bfI_{n-1}\end{bmatrix}\!\otimes\!\begin{bmatrix}\tau_m^2/2\\ \tau_m\end{bmatrix}\right)\!}_{\triangleq\Upsilon_{\bfd}}\bfd.
    \end{align}\label{psid}
\end{subequations}
\subsubsection{Residue Definition}
The augmented measurement vector $\bfZ_k$ is known and depends on the unknown state and measurement noises whose mean values and covariance matrices (CMs) are sought.
To eliminate the state $\bfx_k$, define a non-zero annihilation matrix $\anni(\calO)\in\real^{\aO\times Ln_z}$ of the matrix $\calO$ such that 
\begin{align}
    \anni(\calO)\calO=\bfnul_{\aO\times n_x}, \label{eq:am}
\end{align}
where $\aO\!=\!Ln_z\!-\!\rank(\calO)$ and $\rank(\calO)$ denotes a rank of the matrix $\calO$.
Then, the residue $\tbfZ_k\!\in\!\real^{\aO}$ is defined as\linebreak\\[-7mm]
\begin{align}
	\tbfZ_k=\anni(\calO) \bfZ_k=\calA\calE_{k}+\calB \bfd, \label{mpeMDM}
\end{align}
where $\calA\triangleq \anni(\calO)\!
\begin{bmatrix}\bfGamma,&\bfI_{Ln_z}\end{bmatrix}\in\real^{\aO\times \re}$, $\calB\triangleq\anni(\calO)\bfGamma\Upsilon_{\bfd}\in\real^{\aO\times n_z}$, $\calE_{k}\triangleq\begin{bsmallmatrix}\bfW_k \\ \bfV_k \end{bsmallmatrix}\in\!\real^{\re}$ 
with $\re\!=\!(L-1) n_w\!+\!Ln_v$.

\subsubsection{Frequency Drift Identification}\label{fDriftSec}

Consequently, the mean of the residue $\tbfZ_k$ is a linear function of the unknown frequency drift $\bfd$, i.e.,
\begin{align}
	\!\!\!\!\calR_{\tbfZ}&\triangleq\mean\!\left[\tbfZ_k\right]
    =\calB\bfd.\label{meanTrue}
\end{align}
Although the residue mean $ \calR_{\tbfZ} $ is unknown, 
its sample-based estimate $\widehat{ \calR_{\tbfZ}}= \frac{1}{\bN\!-\!L\!+\!1}\!\sum_{k=0}^{\bN-L}\tbfZ_k $ can be computed from measurements and \eqref{meanTrue} can be modified as
\begin{align}
	\underbrace{\widehat{ \calR_{\tbfZ}}}_{\hspace{-43px}\text{Known vector}\hspace{-20px}}
	=
\overbrace{\calB}^{\hspace{-35px}\text{Known matrix}\hspace{-35px}}
	\underbrace{\bfd}_{\hspace{-20px}\text{Sought drift vector}\hspace{-55px}}
	\ + \
	\overbrace{\!\!\!\bfzeta\!\!\!}^{\hspace{-10px}\text{Unknown \textit{zero-mean} vector}\hspace{-60px}}.\hspace{-10px}\label{meanMeasBeta}
\end{align}
The drift estimates can be obtained, assuming full-rank matrix $\calB$, by the least squares (LS) method from \eqref{meanMeasBeta} as follows
\begin{align}
	\widehat{\bfd} &=\calB^{\dagger}\ \widehat{\calR_{\tbfZ}}\label{estMean}.
\end{align}

The CM of the estimate $\cov\!\big(\widehat{\bfd}\big)$ depends on the resampling period $\tau_m$; a discussion of this is later in Section \ref{influenceTs}. To obtain the resampling period that leads to minimum CM of the estimate, the estimate of the residual $\widehat{\calR_{\tbfZ}}$ and, consequently, the drift $\widehat{\bfd}$ is split into $S\!>\!1$ mutually-uncorrelated\footnote{The residue $\tbfZ_k$ is time-correlated over $L\!-\!1$ time steps. However, the dependencies between individual sub-estimates $\widehat{\bfd^p}$ can be eliminated such that the estimate $\widehat{\calR^p_{\tbfZ}}$ is not calculated from a full period of $M$ residues, but the last $L\!-\!1$ residues are omitted. } 
sub-estimates 
as
\begin{align}
    \widehat{\bfd^p} &=\calB^{\dagger}
    \underbrace{\tfrac{1}{M-L+1}\!\textstyle\sum_{k=0}^{M-L}\tbfZ_{k+(p-1)M}}_{\widehat{\calR^p_{\tbfZ}}}
    \label{estMeanPart}\ ,
\end{align}
for $p=1,2,\ldots,S$ where $M\!=\!\big\lfloor\frac{\bN-L+1}{S}\big\rfloor$.
The sample-based estimate of the covariance of the drift estimate is computed as 
\begin{align}
     \widehat{\cov\!\big(\widehat{\bfd^p}\big)} &= \tfrac{1}{S-1}\!\textstyle\sum_{p=1}^{S}\! \left( \widehat{\bfd^p} -\! \left(\!\tfrac{1}{S}\!\textstyle\sum_{p=1}^{S} \widehat{\bfd^p}\right)\!\!\right)\!\!\Big(\!\cdot\!\Big)\ssT.\label{estcovEstMeanPart}
\end{align}
The estimate of the covariance \eqref{estcovEstMeanPart} can be computed for 
$\te$ different resampling period $\bar{\bftau} \!=\! \left[ \bar{\tau}_{1}, \bar{\tau}_{2}, \dots, \bar{\tau}_{\te}\right] \!\in\! \bbR^{\te}$, where $\bar{\tau}_{p} \!=\! m_{p}\Ts$ with $m_{p}\!\in\!\{1,2,\ldots,\! \big\lfloor \tfrac{N}{L(S+1)-2}\big\rfloor \! \}$,
 and the one for which the covariance is minimal, in a certain sense, can be used to estimate the drift $\widehat{\bfd}$ \eqref{estMean}. Furthermore, the CM of the estimated drift can be estimated from \eqref{estcovEstMeanPart}  using the relation
\begin{align}
     \widehat{\cov\!\big(\widehat{\bfd}\big)} &=\tfrac{\calC(\bN-L+1,L)}{\calC(M-L+1,L)}\widehat{\cov\!\big(\widehat{\bfd^p}\big)},\label{estcovEstMean}
\end{align}
where $\calC(\bN,L) \!=\! \frac{1}{\bN}+\sum_{i=1}^{\min(\bN-1,L-1)} \!\frac{2(\bN-i)}{\bN^2}\frac{L-i}{L}$ is described\linebreak in the Appendix~\ref{sec:appendixB}.

\subsubsection{Relation of Residue and Noises Covariances}
The residue definition \eqref{mpeMDM} leads to the following key properties: 
i)  $\tbfZ_k$ is a \textit{linear} function of the unavailable state and measurement noises (and drift), and 
ii) it can be computed from the measurements.

Consequently, the residue correlation is a linear function of the CMs of the state and measurement noises, i.e.,
\begin{align}
	\calR_{\tbfZ^2}&\triangleq\mean\!\left(\tbfZ_k^{\otimes^2}\right)
    =\calA^{\otimes^2}\calR_{\calE^2} + (\calB\bfd)^{\otimes^2}
    ,\label{noncentMom}
\end{align}
where $\calA^{\otimes^2}\triangleq\calA\otimes\calA$ denotes the Kronecker power, $\calR_{\tbfZ^2}\in\real^{\aO^2}$ is the vectorised residual \textit{correlation} containing the CM elements of the residue $\tbfZ_k$, $\calR_{\calE^2}\in\real^{\re^2}$ contains elements of the CM of the augmented state and measurement noise vector $\calE_{k}$ and will be specified below.
\subsubsection{Noise CM Parametrisation via Structure-Defining Matrices}
The noise covariance $\calR_{\calE^2}$ contains multiple copies of \textit{unique} elements of the noise CMs $\bbfQ$ and $\bfR$, due to either the construction of the residue, application of the Kronecker algebra, or the CM symmetry. For efficient estimation of the noise CMs, it is necessary to express the \textit{unique} elements of the noise CMs $\bbfQ, \bfR$ from the vector $\calR_{\calE^2}$ in \eqref{noncentMom}. An appealing parametrisation, introduced in \cite{Be:74}, defines $\bbfQ, \bfR$ using a set of known (or user-defined) structure-defining matrices $ \bfB_{q_1}^{(i)}\!\in\!\real^{2n\times 2n}$, $ \bfB_{q_2}^{(i)}\!\in\!\real^{2n\times 2n}$ and $ \bfB_r^{(i)}\!\in\!\real^{n_z\times n_z}$ and unknown (and sought) parameters $\bfq_1$, $\bfq_2$, $\bfr$ as
\begin{align}
	\!\!\!\!\bbfQ \!=\! \textstyle\sum_{i=1}^{n}\!q_1^{(i)}\bfB_{q_1}^{(i)}
    \!+\!\sum_{i=1}^{n}\!q_2^{(i)}\bfB_{q_2}^{(i)},\ \ \ \
    \bfR \!=\! \sum_{i=1}^{n_r}r_i\bfB_r^{(i)}\!,\!\!\!\!
\end{align}
where $\bfB_{q_1}^{(i)}$, $\bfB_{q_2}^{(i)}$ and $\bfB_r^{(i)}$ are defined as\!\!\!\!
\begin{subequations}\label{basMat}
\begin{align}
\!\!\!\!\bfB_{q_1}^{(i)}&=
\diag\!\left(\!
\begin{bsmallmatrix}\!
    \!\bfnul_{(i-1)\times 1}\! \\ 1 \\  \!\bfnul_{(n-i)\times 1}\!
\end{bsmallmatrix}\!
\right)\!\!
\begin{bmatrix} 
    \tau_m & 0 \\ 
    0 & 0
\end{bmatrix}, \ i=1,2,\cdots\!, n
\\
\!\!\!\!\bfB_{q_2}^{(i)}&=
\diag\!\left(\!
\begin{bsmallmatrix}\!
    \!\bfnul_{(i-1)\times 1}\! \\ 1 \\ \!\bfnul_{(n-i)\times 1}\!
\end{bsmallmatrix}\!
\right)\!\!
\begin{bmatrix} 
    \tau_m^3/3 & \tau_m^2/2 \\ 
    \tau_m^2/2 & \tau_m 
\end{bmatrix}, \ i=1,2,\ldots\!, n 
\\
\!\!\!\!\bfB_r^{(i)}&= 
\calI(i),  \ i=1,2,\ldots\!, n_r, 
\end{align}
\end{subequations}
where $\calI(j)$ is a function that returns a zero matrix of dimension $n_z \times n_z$ containing ones only at the positions where the $i$-th unique element $r_i$ of the matrix $\bfR$ is located.
For example, for the first unique element $r_1=R_{11}$, the matrix is $\calI(1)\!=\!\diag(\begin{bsmallmatrix}
    \!1 & \bfnul_{1\times (n_z\!-1)\!}
\end{bsmallmatrix})$, and the element $r_{n_z\!+2}\!=\!R_{13}\!=\!R_{31}$, the matrix is $\calI(n_z\!\!+\!1)\!=\!\diag\!\left(\!\begin{bsmallmatrix}\!0 & 0 & 1\! \\ \!0 & 0 & 0\! \\ \!1 & 0 & 0\! \end{bsmallmatrix}\!, \bfnul_{n_z\!-3\times n_z\!-3} \right)\!$.


\subsubsection{Residue Correlation}
Then, the noise CMs parametrisation \eqref{basMat} allows to incorporate a priori knowledge of the noise structure using the \textit{minimum} number of noise CM parameters gathered in the vector $\bftheta_a$. The vectors $\calR_{\calE^2}$ and $\bftheta_a$ are related as follows
\begin{subequations}\label{NpsiM}
\begin{align} 
    \!\!\!\!\calR_{\calE^2}\triangleq
    \mean\!\left(\calE_k^{\otimes^2}\right)
	=&
    \left(\diag\!\left(\bfI_{(L-1)}\!\otimes\!\bbfQ,\ \bfI_{L}\!\otimes\!\bfR\right)\right)_\mathsf{V}\hspace{-2mm} 
    \\
	=& \ \Upsilon_{\!q_1}\bfq_1  +  \Upsilon_{\!q_2}\bfq_2  +  \Upsilon_{\!r}\bfr,
\end{align}
\end{subequations}
where the notation $(\cdot)_\mathsf{V}$ stands for vectorisation of the matrix and the replication matrices $\Upsilon_{\!q_1}\in\real^{\aO^2\times n}$, $\Upsilon_{\!q_2}\in\real^{\aO^2\times n}$, and $\Upsilon_{\!r}\in\real^{\aO^2\times n_r}$ are defined as
\begin{subequations}\label{Npsi}
\begin{align} 
    \Upsilon_{\!q_1}&\triangleq
        \bigg[\!
		\begin{smallmatrix}\!      
            \begin{bsmallmatrix}
				\bfI_{(L-1)}\otimes\bfB_{q_1}^{(1)} & \bfnul\\
				\bfnul & \bfnul
			\end{bsmallmatrix}_\mathsf{V},
			& \cdots \ , 
            &
			\begin{bsmallmatrix}
				\bfI_{(L-1)}\otimes\bfB_{q_1}^{(n)} & \bfnul\\
				\bfnul & \bfnul
			\end{bsmallmatrix}_\mathsf{V}
		\end{smallmatrix}\!\bigg],
    \\[-0mm]
    \Upsilon_{\!q_2}&\triangleq
        \bigg[\!
		\begin{smallmatrix}\!      
            \begin{bsmallmatrix}
				\bfI_{(L-1)}\otimes\bfB_{q_2}^{(1)} & \bfnul\\
				\bfnul & \bfnul
			\end{bsmallmatrix}_\mathsf{V},
			& \cdots \ , 
            &
			\begin{bsmallmatrix}
				\bfI_{(L-1)}\otimes\bfB_{q_2}^{(n)} & \bfnul\\
				\bfnul & \bfnul
			\end{bsmallmatrix}_\mathsf{V}
		\end{smallmatrix}\!\bigg],
    \\[-0mm]
    \Upsilon_{\!r} &\triangleq 
    \bigg[\!
		\begin{smallmatrix}\!      
            \begin{bsmallmatrix}
				\bfnul & \bfnul\\
				\bfnul & \bfI_{L}\otimes\bfB_{r}^{(1)}
			\end{bsmallmatrix}_\mathsf{V},
			& \cdots \ ,
            &
            \begin{bsmallmatrix}
				  \bfnul & \bfnul\\
			    \bfnul & \bfI_{L}\otimes\bfB_{r}^{(n_r)}
			\end{bsmallmatrix}_\mathsf{V}
		\end{smallmatrix}\!\bigg].
\end{align}
\end{subequations}

\subsubsection{Residue Correlation Unification}
Similarly to the noise covariance $\calR_{\calE^2}$ in \eqref{noncentMom}, the residue correlation $\calR_{\tbfZ^2}$ also contains redundant copies of the same elements. Therefore, unique elements of the vector $\calR_{\tbfZ^2}$ gathered in the vector denoted as $\calR_{\tbfZ^2}^\rmU\in\real^{\aO(\aO+1)/ 2}$ are selected using a unification matrix denoted as $\Xi_{\tbfZ^2}\in\real^{ (\aO(\aO+1)/ 2) \times \aO^2 }$. Therefore, the following relationship holds
\begin{align}
	\calR_{\tbfZ^2}^{\rmU}=\Xi_{\tbfZ^2}\calR_{\tbfZ^2}.\label{NXiM}
\end{align}

\subsubsection{Noise Parameter Estimation}
Consequently, the correlation of the residue is a linear function of the noise parameters and equations \eqref{NpsiM} and the \textit{kron}-square of the state noise mean vector (drift) \eqref{noncentMom}, i.e.,
\begin{align}
	\!\!\!\!\calR_{\tbfZ^2}
	\!&=
    \Xi_{\tbfZ^2}\!
    \begin{bmatrix}
        \calA^{\otimes^2}\!\!
        \begin{bmatrix}
            \Upsilon_{\!q_1},&\!\! 
            \Upsilon_{\!q_2},&\!\! 
            \Upsilon_{\!r}
        \end{bmatrix}
        \!\!,&\!\calB^{\otimes^2}\!  
    \end{bmatrix}
    \!\!
    \begin{bsmallmatrix}
        \bfq_1 \\ \bfq_2 \\ \bfr \\ \bfd^{\otimes^2}
    \end{bsmallmatrix}.
    \!\!\!\label{covTrue}
\end{align}
The residue correlation $ \calR_{\tbfZ^2} $ \eqref{NXiM} is unknown. However, its sample-based estimate $\widehat{ \calR_{\tbfZ^2}^{\rmU}}\!=\!\Xi_{\tbfZ^2}\widehat{ \calR_{\tbfZ^2}}\!=\! \frac{\Xi_{\tbfZ^2}}{\bN\!-\!L\!+\!1}\!\sum_{k=0}^{\bN-L}\tbfZ_k^{\otimes^2} $ can be computed and \eqref{covTrue} can be modified as
\begin{align}
	\!\!\!\!\underbrace{\widehat{ \calR_{\tbfZ^2}^\rmU}\!}_{\hspace{-43px}\text{Known vector}\hspace{-43px}}
	=
   \underbrace{\Xi_{\tbfZ^2}\!\begin{bmatrix}
        \calA^{\otimes^2}\!\!
        \begin{bmatrix}
        \Upsilon_{\!q_1},&\!\!\! 
        \Upsilon_{\!q_2},&\!\!\! 
        \Upsilon_{\!r}
        \end{bmatrix}
        \!\!,&\!\calB^{\otimes^2}\!  
    \end{bmatrix}\!}_{\hspace{-20px}\text{Known matrix}\hspace{-20px}}
	\overbrace{\!\begin{bsmallmatrix}
        \bfq_1 \\ \bfq_2 \\ \bfr \\ \bfd^{\otimes^2}
    \end{bsmallmatrix}\!}^{\hspace{-25px}\text{Sought noise parameters}\hspace{-25px}}
	+
	\underbrace{\!\!\!\!\bfnu\!\!\!\!\!}_{\hspace{-70px}
    \begin{smallmatrix}
	    \text{Unknown} \\
        \text{\textit{zero-mean}} \\
        \text{vector}
	\end{smallmatrix}\hspace{-70px}}\!.\hspace{-8px}\label{CovParEst}
\end{align}

Analogously to the drift $\bfd$ and its estimate\eqref{estMean}, the noise parameters and squared drift can directly be estimated by the LS method from \eqref{CovParEst}. However, in this straightforward solution the MSE of a subset of estimated parameter will be significantly higher than the MSE of the remaining parameters due to the ``uneven'' influence of the sampling period, that is analyzed Section \ref{influenceTs}. To equalize the MSE for all estimated parameters, we, therefore, propose a \textit{two-step} procedure for sequential noise parameters estimation based on two purposely selected (down)-sampled periods.



The \textit{first} step, it can be shown that for a small resampling period, e.g., $\tau_m \leq 100s$, the influence of the noise parameters $\bfq_2$ and $\bfd$ is negligible, and the equation \eqref{CovParEst} can be considered in the following form
\begin{align}
	\!\!\!\!\widehat{ \calR_{\tbfZ^2}^\rmU}
	\approx
   \Xi_{\tbfZ^2}
    \calA^{\otimes^2}\!\!
    \begin{bmatrix}
    \Upsilon_{\!q_1}\!,&\!\! 
    \Upsilon_{\!r}
    \end{bmatrix}\!\! 
	\begin{bsmallmatrix}
        \!\bfq_1\! \\[1mm] \!\bfr\!
    \end{bsmallmatrix}
	\!+
	\bfnu.\label{CovParEst_q1r}
\end{align}
The noise parameters $\bfq_1$ and $\bfr$ estimates can be obtained by LS from \eqref{CovParEst_q1r} as follows
\begin{align}
	\!\!\!\begin{bsmallmatrix}\!\widehat{\bfq_1}\! \\ \!\widehat{\bfr}\! \end{bsmallmatrix} \!
    &=\! \left(\Xi_{\tbfZ^2}\calA^{\otimes^2}\!\begin{bmatrix}
        \Upsilon_{\!q_1}\!,&\!\! \Upsilon_{\!r}   
    \end{bmatrix}\right)^{\!\dagger}\! \widehat{ \calR_{\tbfZ^2}^\rmU}\label{Est_CovParEst_q1r}.
\end{align}

The \textit{second} step, the influence of the noise parameters $\bfq_1$, $\bfr$, and $\bfd$ can be eliminated from equation \eqref{CovParEst} by using the available estimates $\widehat{\bfq_1}, \widehat{\bfr}$ \eqref{Est_CovParEst_q1r} and $\widehat{\bfd}$ \eqref{estMean}, and subsequently estimate the last remaining parameter $\bfq_2$ as follows
\begin{align}
	\!\!\!\!\widehat{\bfq_2} \!&=\!\! \left(\Xi_{\tbfZ^2}\calA^{\otimes^2}\!\Upsilon_{\!q_2}\!\right)^{\!\!\dagger}\!\!
    \bigg(\!\widehat{ \calR_{\tbfZ^2}^\rmU}\!-\! \Xi_{\tbfZ^2}\!\!
    \begin{bmatrix}
       \calA^{\otimes^2}\!\! 
       \begin{bmatrix}
       \Upsilon_{\!q_1},&\!\!
       \Upsilon_{\!r}\!
       \end{bmatrix}\!\!,&\!
       \calB^{\otimes^2}\!
    \end{bmatrix}\!\!\!
    \begin{bsmallmatrix}
       \widehat{\bfq_1} \\ \widehat{\bfr} \\ \!\widehat{\bfd}^{\otimes^2}\!\!
    \end{bsmallmatrix}\!\bigg)\!.\!\!\!\!\label{Est_CovParEst_q2}
\end{align}
The CM of the estimates $\widehat{\bfq_1}$, $\widehat{\bfr}$, and $\widehat{\bfq_2}$ depends on the resampling period $\tau_m$. The technique for obtaining the resampling period $\tau_m$ that minimizes the CM of the estimates (similar to the estimation of the drift \eqref{estMean}) is to use $S$ mutually uncorrelated sub-estimates 
as
\begin{subequations}\label{Est_CovParEst_q12rPart}
\begin{align}
    \!\!\!\!\!\!\begin{bsmallmatrix}\!\widehat{\bfq_1^p}\! \\ \!\widehat{\bfr^p}\! \end{bsmallmatrix} \!\!
    &=\!\! \left(\Xi_{\tbfZ^2}\calA^{\otimes^2}\!\begin{bmatrix}
        \Upsilon_{\!q_1},&\!\! \Upsilon_{\!r}   
    \end{bmatrix}\right)^{\!\dagger}\! \widehat{\calR^{p,\rmU}_{\tbfZ^2}},\!\!\!\!\label{Est_CovParEst_q1rPart}
    \\
	\!\!\!\!\widehat{\bfq_2^p} \!&=\!\! \left(\Xi_{\tbfZ^2}\calA^{\otimes^2}\!\Upsilon_{\!q_2}\!\right)^{\!\!\dagger}\!
    \!\bigg(\!\widehat{ \calR_{\tbfZ^2}^{p,\rmU}}\!\!
    -\! \Xi_{\tbfZ^2}\!\!
    \begin{bmatrix}
       \!\calA^{\otimes^2}\!\!
       \begin{bmatrix}
       \Upsilon_{\!q_1}\!,&\!\!
       \Upsilon_{\!r}\!
       \end{bmatrix}\!\!,&\!
       \calB^{\otimes^{\!2}}\!
    \end{bmatrix}\!\!\!
    \begin{bsmallmatrix}
       \widehat{\bfq_1} \\ \widehat{\bfr} \\ \!\widehat{\bfd}^{\otimes^2}\!\!\!
    \end{bsmallmatrix}\!\bigg)\!\!\!\!\!\label{Est_CovParEst_q2Part}
\end{align}
\end{subequations}
for $p\!=\!1,2,\!\ldots\!,S$, where $\widehat{\calR^{p,\rmU}_{\tbfZ^2}}\!=\!\!\Xi_{\tbfZ^2}\widehat{\calR^{p}_{\tbfZ^2}}$, $\widehat{\calR^{p}_{\tbfZ^2}} = \tfrac{1}{M-L+1}\!\sum_{k=0}^{M-L}\tbfZ_{k+(p-1)M}^{\otimes^2}$ and $M\!=\!\!\left\lfloor\frac{\bN-L+1}{S}\right\rfloor$.

Using the sub-estimates \eqref{Est_CovParEst_q12rPart}, the appropriate resampling period can be found for the estimate of the pair $\begin{bsmallmatrix}\!\widehat{\bfq_1}\! \\ \!\widehat{\bfr}\!\end{bsmallmatrix}$ \eqref{Est_CovParEst_q1r} as well as for the estimate of $\widehat{\bfq_2}$ \eqref{Est_CovParEst_q2}. As a consequence, each of the chosen resampling period minimizes the CM of the respective estimate error, 
which can be calculated as 
\begin{subequations}\label{estcovEst_CovParEst_q1rOrq2}
\begin{align}
     \widehat{\cov\!\Big(\!\begin{bsmallmatrix}\!\widehat{\bfq_1}\! \\ \!\widehat{\bfr}\!\end{bsmallmatrix}\!\Big)}&=\tfrac{\calC(\bN-L+1,L)}{\calC(M-L+1,L)} \ \widehat{\cov\!\Big(\!\begin{bsmallmatrix}\!\widehat{\bfq_1^p}\! \\ \!\widehat{\bfr^p}\!\end{bsmallmatrix}\!\Big)},\label{estcovEst_CovParEst_q1r}
     \\
     \widehat{\cov\!\big(\widehat{\bfq_2} \!\big)}&=\tfrac{\calC(\bN-L+1,L)}{\calC(M-L+1,L)} \ \widehat{\cov\!\big(\widehat{\bfq_2^p}\big)},\label{estcovEst_CovParEst_q2}
\end{align} 
\end{subequations}
where 
\begin{subequations}\label{estcovEst_CovParEst_q1rOrq2Part}
\begin{align}
     \!\!\!\widehat{\cov\!\Big(\!\begin{bsmallmatrix}\widehat{\bfq_1^p} \\ \widehat{\bfr^p}\end{bsmallmatrix}\!\Big)} 
     \!\!&=\! \tfrac{1}{S-1}\!\textstyle\sum_{p=1}^{S}\! \left( \begin{bsmallmatrix}\widehat{\bfq_1^p} \\ \widehat{\bfr^p}\end{bsmallmatrix} \!-\! \left(\!\frac{1}{S}\!\textstyle\sum_{p=1}^{S} \begin{bsmallmatrix}\widehat{\bfq_1^p} \\ \widehat{\bfr^p}\end{bsmallmatrix}\right)\!\!\right)\!\!\Big(\!\cdot\!\Big)\sT\!,\!\!\!\label{estcovEst_CovParEst_q1rPart}
     \\
     \!\!\!\widehat{\cov\!\big(\widehat{\bfq_2^p} \big)} \!\!&=\! \tfrac{1}{S-1}\!\textstyle\sum_{p=1}^{S}\! \left( \widehat{\bfq_2^p} \!-\! \left(\!\tfrac{1}{S}\!\textstyle\sum_{p=1}^{S} \widehat{\bfq_2^p} \right)\!\!\right)\!\!\Big(\!\cdot\!\Big)\sT\!.\!\!\!\label{estcovEst_CovParEst_q2Part}
\end{align}
\end{subequations}

\subsubsection{Influence of Resampling Period on Noise Parameter Estimation, Notes, and Algorithm}\label{influenceTs}
Let's focus on \textit{observations} that determined why individual estimates are calculated in these particular ways. In this section consider a set of AHM clocks as described in Section \ref{sec:SimSce}. The \textit{observations} presented below are illustrative; however, they have been valid for various model parameters as well.

\begin{figure}[H]
    \centering
    \vspace*{-3mm}
    \includegraphics[width=0.48\textwidth]{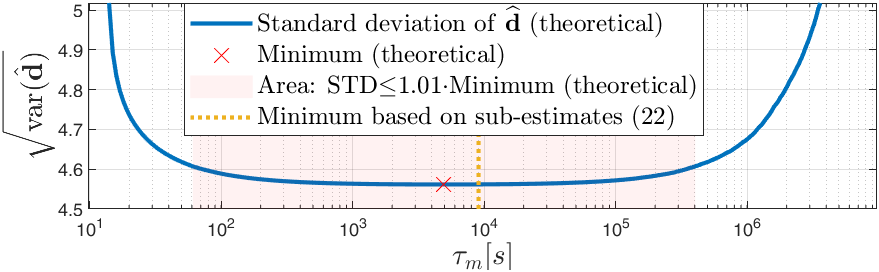}
    \vspace*{-3mm}
    \caption{The influence of $\tau_m$ on the STD of the mean estimate $\widehat{\bfd}$.}
    \vspace*{-3mm}
    \label{fig:dEst}
\end{figure}

\hspace{-3.5mm}\textit{Observation I:} To estimate $\bfd$ \eqref{estMean}, the choice of the resampling period $\tau_m$ is crucial, as it can significantly affect the variance of the estimate, as shown in Fig. \ref{fig:dEst}. However, it can be seen the wide range between $\tau_m\!=\!\num{60}$\si{\second} and $\tau_m\!=\!\num{41000}$\si{\second} where the STD of the estimate differs from the minimum by 1\%.

\hspace{-3.5mm}\textit{Observation II:}. 
The appropriate resampling periods $\tau_m$ determined based on sub-estimates by \eqref{estcovEstMeanPart} and \eqref {estcovEst_CovParEst_q1rOrq2Part} are very close to the resampling periods that lead to the minimum variance of the estimate calculated both theoretically and using Monte Carlo (MC) simulations i.e., $\tau_m=\num{4800}$\si{\second}. This can be seen in Fig.~\ref{fig:dEst} for the frequency drift identification.

\begin{figure}[H]
    \centering
    \vspace*{-3mm} 
    \includegraphics[width=0.48\textwidth]{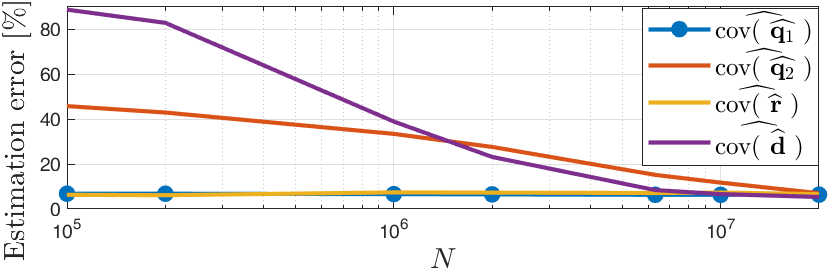}
    \vspace{-3mm}
    \caption{The estimation error of the CM of the noise parameter estimate.}
    \vspace*{-3mm}
    \label{fig:error_cov_est}
\end{figure}

\hspace{-3.5mm}\textit{Observation III:} The error of the estimated CM of the estimate based on \eqref{estcovEstMean} and \eqref{estcovEst_CovParEst_q1rOrq2} (compared to the CM estimated based on MC simulations) for different value of measurements $N$ (with $\Ts=5s$) is shown in Fig.~\ref{fig:error_cov_est}.
The average error of CMs for $\widehat{\bfq_1}$ and $\widehat{\bfr}$ is 6\%, independently of the total number of measurements $N$ with $\Ts=5s$. In contrast, the error of CMs for $\widehat{\bfq_2}$ and $\widehat{\bfd}$ decreases as the value of $N$ increases.

\begin{figure}[H]
    \centering
    \vspace*{-3mm} 
    \includegraphics[width=0.48\textwidth]{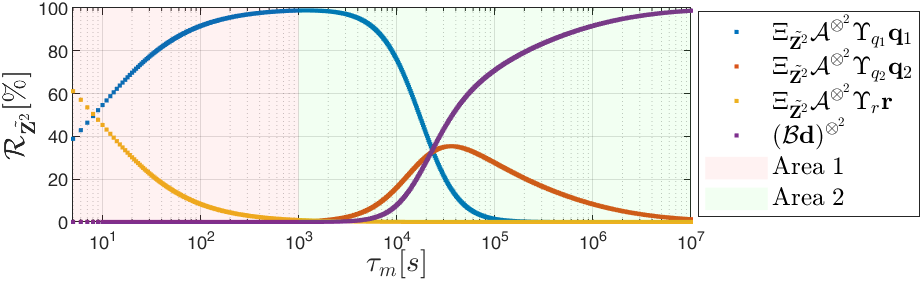}
    \vspace{-3mm}
    \caption{The influence of $\bfq_1$, $\bfq_2$, $\bfr$, and $\bfd$ to the correlation value $\calR_{\tbfZ^2}$}
    \vspace*{-3mm}
    \label{fig:XiRes2}
\end{figure}

\hspace{-3.5mm}\textit{Observation IV:} The resampling time $\tau_m$ determines the ratio of the contributions of the unknown parameters $\bfq_1$, $\bfq_2$, $\bfr$, and $\bfd$ \eqref{covTrue} to the correlation value $\calR_{\tbfZ^2}$, as shown in Fig.~\ref{fig:XiRes2}.
\hspace{-3.5mm}Therefore, for a small resampling period $\tau_m$ (Area 1), the correlation $\calR_{\tbfZ^2}$ is formed exclusively by the parameters $\bfq_1$ and $\bfr$. As a result, the parameters $\bfq_1$ and $\bfr$ can be estimated by \eqref{Est_CovParEst_q1r} individually, without knowledge of the remaining parameters $\bfq_2$ and $\bfd$.
For longer resampling period $\tau_m$  (Area 2), the correlation $\calR_{\tbfZ^2}$ is composed of all parameters. However, since the estimates $\bfd$ \eqref{estMean} and $\bfq_1, \bfr$ \eqref{Est_CovParEst_q1r} are already available, their effects can be removed from the equation, and the remaining parameter $\bfq_2$ can be estimated by \eqref{Est_CovParEst_q2}.

\hspace{-3.5mm}\textit{Observation V:}
The appropriate resampling period for both $\bfq_1, \bfr$ parameters and $\bfq_2$ parameter yields the minimum variance of these estimates (see Observation II), corresponds to the resampling period at which the parameters have a significant contributions (see Observation IV) i.e., $\tau_m\!=\!8$\si{\second} for $\bfq_1, \bfr$ and $\tau_m\!=\!\num{36000}$\si{\second} for $\bfq_2$. 


\hspace{-3.5mm}\textit{Note I:} The noise can be arbitrarily distributed, and the assumption of Gaussian distributed noises are not necessary.

\begin{figure}[H]
    \centering
    \vspace*{-3mm} 
    \includegraphics[width=0.48\textwidth]{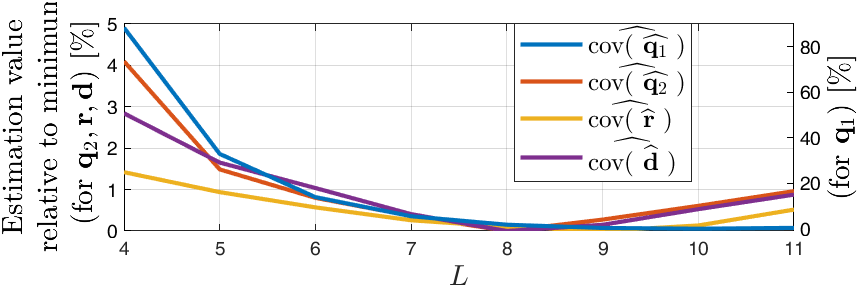}
    \vspace{-3mm}
    \caption{The value of the estimated CM of the noise parameter estimate.}
    \vspace*{-3mm}
    \label{fig:cov_est_vs_L}
\end{figure}

\hspace{-3.5mm}\textit{Note II:} The technique for finding the appropriate resampling period from set $\bar{\bftau}$ can also be applied to 
$\ell$ different values of $L$ i.e., $\bar{\bfL} \!=\! \left[ \bar{L}_{1}, \bar{L}_{2}, \dots, \bar{L}_{\ell}\right] \!\in\! \bbR^{\ell}$, where $\bar{L}_{p} \!\in\! \bbN$,
to find the appropriate one for which the noise parameters (as a whole) have minimum variance. The appropriate value is $L=8$ as is shown in Fig.~\ref{fig:cov_est_vs_L}. 


\begin{table}[H]
\vspace{-5mm}
    \centering
        \begin{algorithm}[H] 
        \caption{The MDM estimation algorithm} \label{alg:MDM}
            \begin{algorithmic}[1]
                \Require $\bfz_{1:N}$, $\bar{\bftau}$, $\bar{\bfL}$, $S$
                    \For{$L \in \bar{\bfL}$}
                        \For{$\tau_m \in \bar{\bftau}$}
                             \State Compute residue $\tbfZ_k$ \eqref{mpeMDM}.
                             \State Estimate $\widehat{\bfd^p}$ as in \eqref{estMeanPart} 
                             and $\begin{bsmallmatrix}\!\widehat{\bfq_1^p}\! \\ \!\widehat{\bfr^p}\! \end{bsmallmatrix}$ \eqref{Est_CovParEst_q1rPart} 
                             \State Estimate $\widehat{\cov\!\big(\widehat{\bfd}\big)}$ as in \eqref{estcovEstMean}
                             and $\widehat{\cov\!\Big(\!\begin{bsmallmatrix}\!\widehat{\bfq_1}\! \\ \!\widehat{\bfr}\!\end{bsmallmatrix}\!\Big)}$  \eqref{estcovEst_CovParEst_q1r}.
                        \EndFor
                        \State Estimate $\widehat{\bfd}$ \eqref{estMean} for such a $\tau_m$ that $\widehat{\cov\!\big(\widehat{\bfd}\big)}$ is minimal.
                        \State Estimate $\begin{bsmallmatrix}\!\widehat{\bfq_1}\! \\ \!\widehat{\bfr}\!\end{bsmallmatrix}$ \eqref{Est_CovParEst_q1r} for such a $\tau_m$ that $\widehat{\cov\!\Big(\!\begin{bsmallmatrix}\!\widehat{\bfq_1}\! \\ \!\widehat{\bfr}\!\end{bsmallmatrix}\!\Big)}$ is minimal.
                        \For{$\tau_m \in \bar{\bftau}$}
                             \State Estimate $\widehat{\bfq_2^p}$ \eqref{Est_CovParEst_q2Part} with $\widehat{\bfq_1}$, $\widehat{\bfr}$, and $\widehat{\bfd}$ from steps 7-8. 
                             \State Estimate $\widehat{\cov\!\big(\widehat{\bfq_2}\big)}$ \eqref{estcovEst_CovParEst_q2}. 
                        \EndFor
                        \State Estimate $\widehat{\bfq_2}$ \eqref{Est_CovParEst_q2} for such a $\tau_m$ that $\widehat{\cov\!\big(\widehat{\bfq_2}\big)}$ is minimal.
                    \EndFor
                    \State Through all $L$, pick estimates $\begin{bsmallmatrix}\!\widehat{\bfq_1}\! \\ \!\widehat{\bfr}\!\end{bsmallmatrix}$, $\widehat{\bfd}$, and $\widehat{\bfq_2}$ from steps 7, 8, and 13, for which their CMs are minimal.
                \Ensure Estimated parameters $\hat{\bftheta}$ and their CMs.
            \end{algorithmic}
        \end{algorithm}
\vspace{-6mm}     
\end{table}

\section{Results} \label{sec:Results}
This section presents illustration and comparison of both identification methods for simulated and real data.
\subsection{Simulated data}\label{sec:SimSce}
The ensemble consists of four active hydrogen maser (AHM) clocks with true parameters listed in Table~\ref{Tab:Synth1} and Table~\ref{Tab:Synth2}. Other parameters considered are: 
$\Ts=\qty{5}{\second}$, $N=\num{6.312e6}$ (i.e., this corresponds to a one-year data period).
The parameter for HCOV is $\te = 20$, where averaging times $\bar{\tau}_{p}$ are evenly spaced in a log-space ranging \qty{5}{\second} to  \qty{3.15e7}{\second}
and 
the parameters for MDM are $S\!=\!100$, $\te \!=\! 400$, where resampling period $\bar{\tau}_{p}$ are evenly spaced in a log-space ranging from \qty{5}{\second} to $5\!\cdot\!\!\big\lfloor \tfrac{\num{6312000}}{101\cdot L-2}\big\rfloor $\si{\second}, and $\bar{\bfL}\!=\! \left[ 4, 5, 6, 7, 8, 9, 10, 11 \right]$ i.e., $\ell\!=\!8$. 


\begin{figure*}[!ht]
    \centering
    \includegraphics[width=0.85\textwidth]{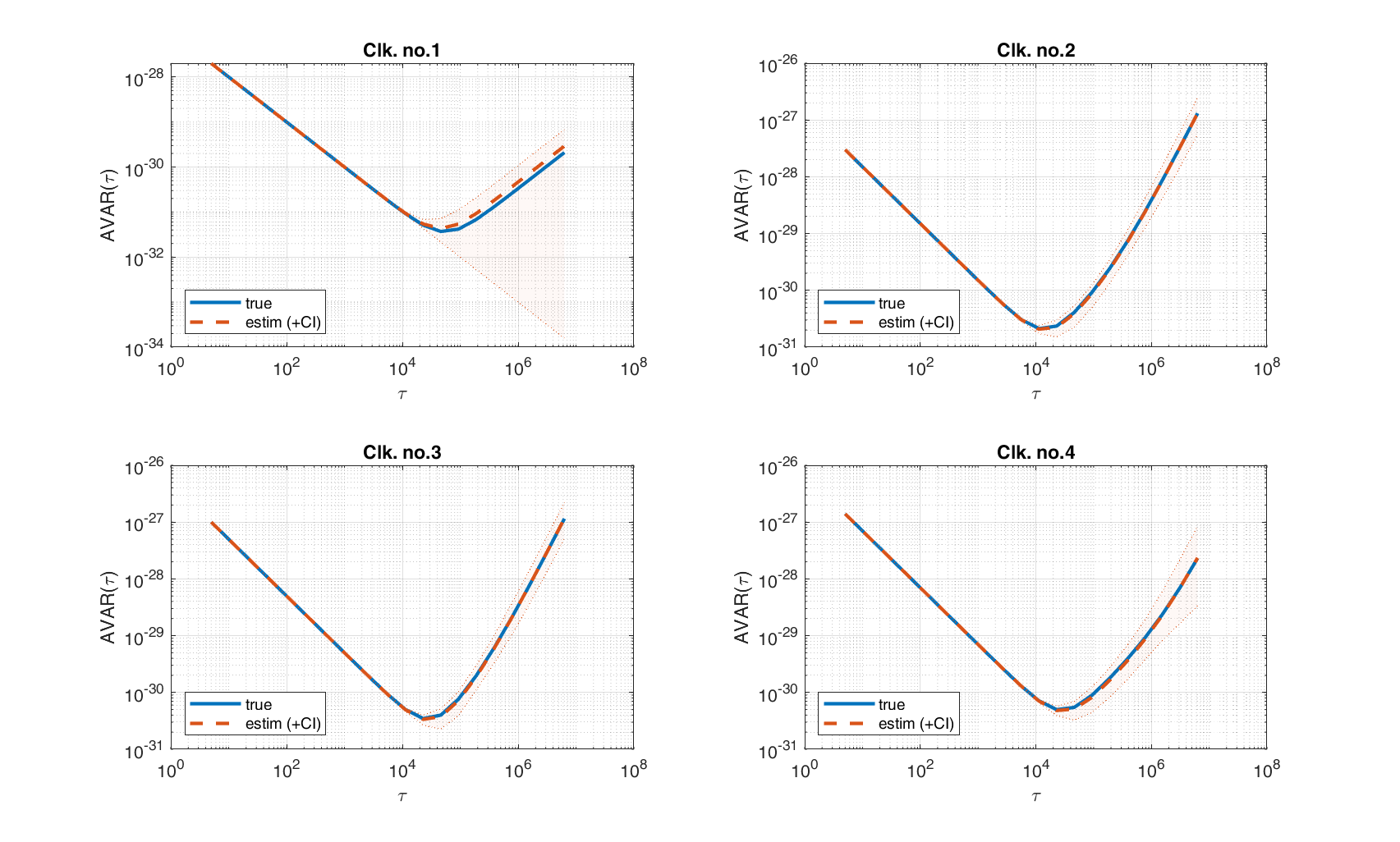}
    \caption{Clocks AVAR for true and estimated parameters (HCOV method - simulation).}
    \label{fig:clkHCOV}
\end{figure*}

\begin{figure*}[!ht]
    \centering
    \includegraphics[width=0.85\textwidth]{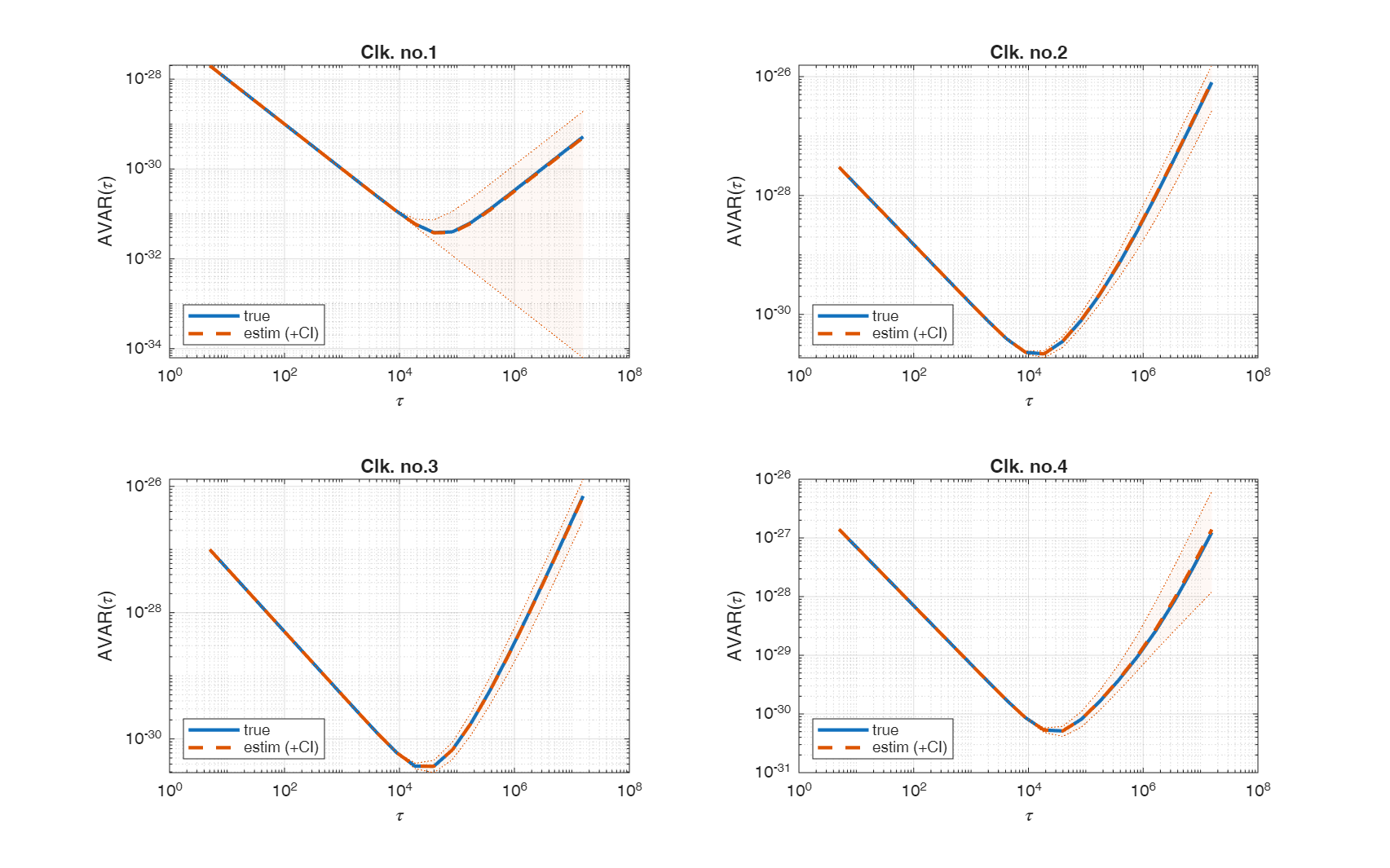}
    \caption{Clocks AVAR for true and estimated parameters (MDM - simulation).}
    \label{fig:clkMDM}
\end{figure*}



The resulting average and standard deviations of the estimates for both methods (HCOV and MDM) based on 100~Monte Carlo simulations are summarized in 
Tables~\ref{Tab:Synth1}~and~\ref{Tab:Synth2}. Both methods seems to provide noise parameter estimates of almost identical quality. A more significant difference can be observed in the uncertainty estimation for parameter $q_2$, where the HCOV method provides optimistic CM, whereas MDM provided CM corresponding to those based on MC simulations. The graphical representation of the results is provided in  Fig.~\ref{fig:clkHCOV}~and~\ref{fig:clkMDM}. The figures present the Allan variance (AVAR) of the individual clocks computed analytically using the true parameters (blue line) and using the average values of the estimated parameters from MC simulations (dashed red line). The confidence intervals (CIs) of the AVAR are shown as shaded areas. The AVARs computed using MC average of parameter estimates are close to the AVARs computed using true parameters. In both cases, a wide confidence interval is obtained for the pivot clock for averaging times greater then \qty{1E4}{\second}. This behavior can be attributed to (i) the assumption of zero drift for the pivot clock, (ii) the low value of the parameter $q_2$ for the pivot clock, and (iii) the limited length of the measurement record. Although the width of a confidence intervals could be further decreased by increasing the duration of measurement, a substantial reduction would require measurement duration on the order of years.

\begin{table*}[!ht]
    \caption{Specification of ensemble clocks parameters and an overview of their estimates.}    
    \centering
    \vspace{-2mm}
    \begin{tblr}{
        colspec={c|c|cccc|cccc|ccc},
        cell{1}{3,7} = {c=4}{c},
        cell{1}{11} = {c=3}{c}
        }
         & & $q_1^{(i)}$ $(\times \num{1e-27})$  & & & & $q_2^{(i)}$ $(\times \num{1e-35})$ & & & & $d^{(i)}$ $(\times \num{1e-21})$ & & &\\
         & $i$ & 1 & 2 & 3 & 4 & 1 & 2 & 3 & 4 & 2 & 3 & 4 \\
        \hline \hline         
         & true & 1 & 1.5 & 5 & 7 & 0.1 & 2 & 1.5 & 2.5 &  8 & 7.5 & 3 \\
         \hline 
        \SetCell[r=3]{c}  \rotatebox{90}{HCOV} & MC-Based $\mean\! \big(\htheta\big)$ & 1 & 1.5 & 5.001 & 7 & 0.133& 1.869& 1.328& 2.197&  7.982 & 7.414 & 3.051 \\[-1.1mm]         
         & MC-Based $\std\! \big( \htheta \big)$ & 0.003 & 0.004& 0.006& 0.008& 0.048& 0.251& 0.206& 0.448&  0.761 & 0.810 & 0.916  \\[-1.1mm]      
         & Method provided  $\std\! \big(\htheta \big)$  
         & 0.002 & 0.004& 0.007& 0.009& 0.043& 0.093& 0.092& 0.131& 0.815 & 0.696 & 0.876  \\[-0.5mm]     
         \hline
         \SetCell[r=3]{c} \rotatebox{90}{MDM}& MC-Based $\mean\! \big(\htheta\big)$
         & 1 & 1.5 & 4.999 & 7 & 0.089 & 2.012 & 1.495 & 2.488 & 8.137 & 7.431 & 3.187 \\[-1.1mm]         
         & MC-Based $\std\! \big( \htheta \big)$ 
         & 0.004 & 0.004 & 0.008 & 0.011 & 0.070 & 0.115 & 0.126 & 0.199 & 0.902 & 0.688 & 0.991  \\[-1.1mm]       
          & Method provided  $\std\! \big(\htheta \big)$  
         & 0.004 & 0.004 & 0.008 & 0.011 & 0.077 & 0.153 & 0.151 & 0.199 & 0.842 & 0.755 & 0.944  \\       
    \end{tblr}
    \label{Tab:Synth1}
\end{table*} 



\begin{table*}[!ht]
    \caption{Specification of the CM $\bfR$ coefficients and an overview of their estimates.}
    \centering
    \vspace{-2mm}
    \begin{tblr}{
        colspec={c|c|cccccc},
        }
         & $(\times \num{1e-27})$ & $R_{11}$ &$R_{12}$ & $R_{13}$& $R_{22}$& $R_{23}$& $R_{33}$\\
        \hline \hline         
         & true 
         & 9 & 6 & 5 & 8.7 & 4 & 9.5  \\
         \hline
        \SetCell[r=3]{c}  \rotatebox{90}{HCOV} & MC-Based $\mean\! \big(\htheta\big)$ 
        & 8.999 & 5.999 & 4.999 & 8.698 & 3.999 & 9.497   \\[-1.1mm]         
         & MC-Based $\std\! \big(\htheta\big)$ 
         & 0.013  & 0.011 & 0.011 & 0.018 & 0.013 & 0.025   \\[-1.1mm]      
          & Method provided  $\std\! \big(\htheta \big)$ 
          & 0.01  & 0.008 & 0.008 & 0.016 & 0.009 & 0.020   \\[-0.5mm]      
         \hline
         \SetCell[r=3]{c} \rotatebox{90}{MDM} & MC-Based $\mean\! \big(\htheta\big)$ 
         & 9.001 & 6.001& 4.999 & 8.703 & 4 & 9.500  \\[-1.1mm]       
         & MC-Based $\std\! \big(\htheta\big)$ 
         & 0.014 & 0.013 & 0.012 & 0.023 & 0.010 & 0.028    \\[-1.1mm]        
          & Method provided  $\std\! \big(\htheta \big)$  
         & 0.013 & 0.012 & 0.013 & 0.021 & 0.013 & 0.027
    \end{tblr}
    \label{Tab:Synth2}
\end{table*}  

\begin{figure*}[!ht]
    \centering
    \includegraphics[width=0.85\textwidth]{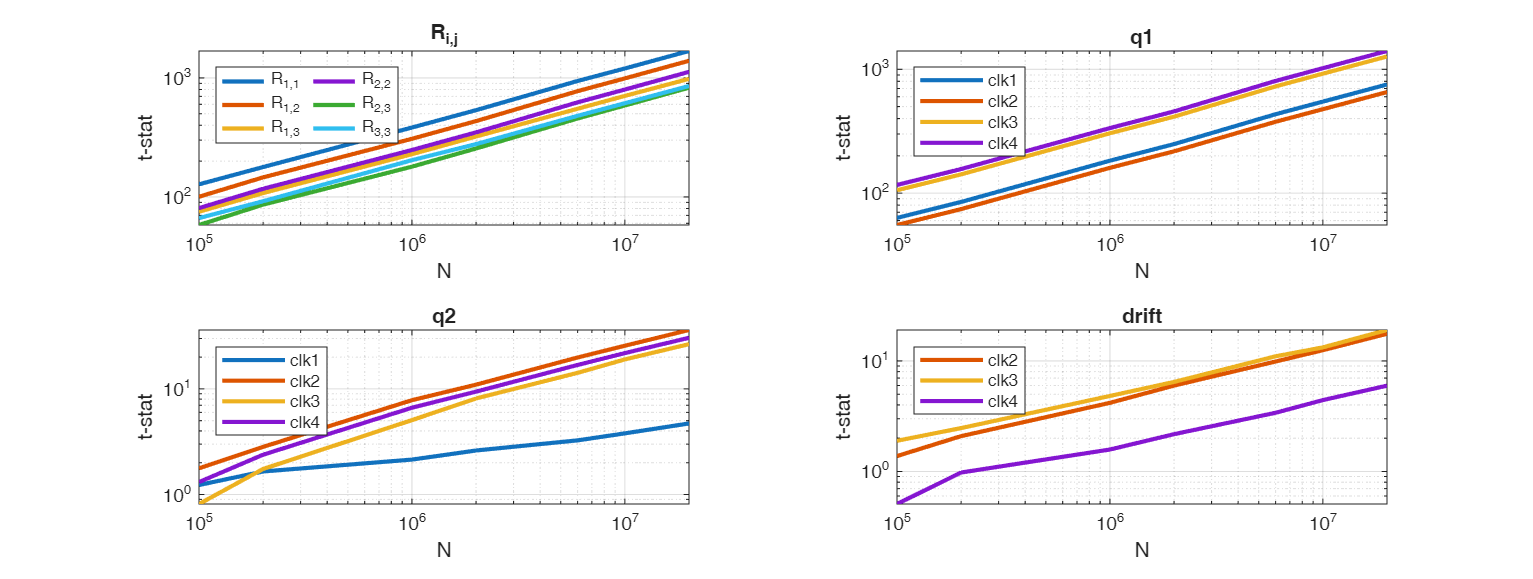}
    \caption{HCOV method: quality of the noise parameter estimates for different values of $N$ ($\Ts =  \qty{5}{\second}$).}
    \label{fig:tstatHCOV}
\end{figure*}
\begin{figure*}[!ht]
    \centering
    \includegraphics[width=0.85\textwidth]{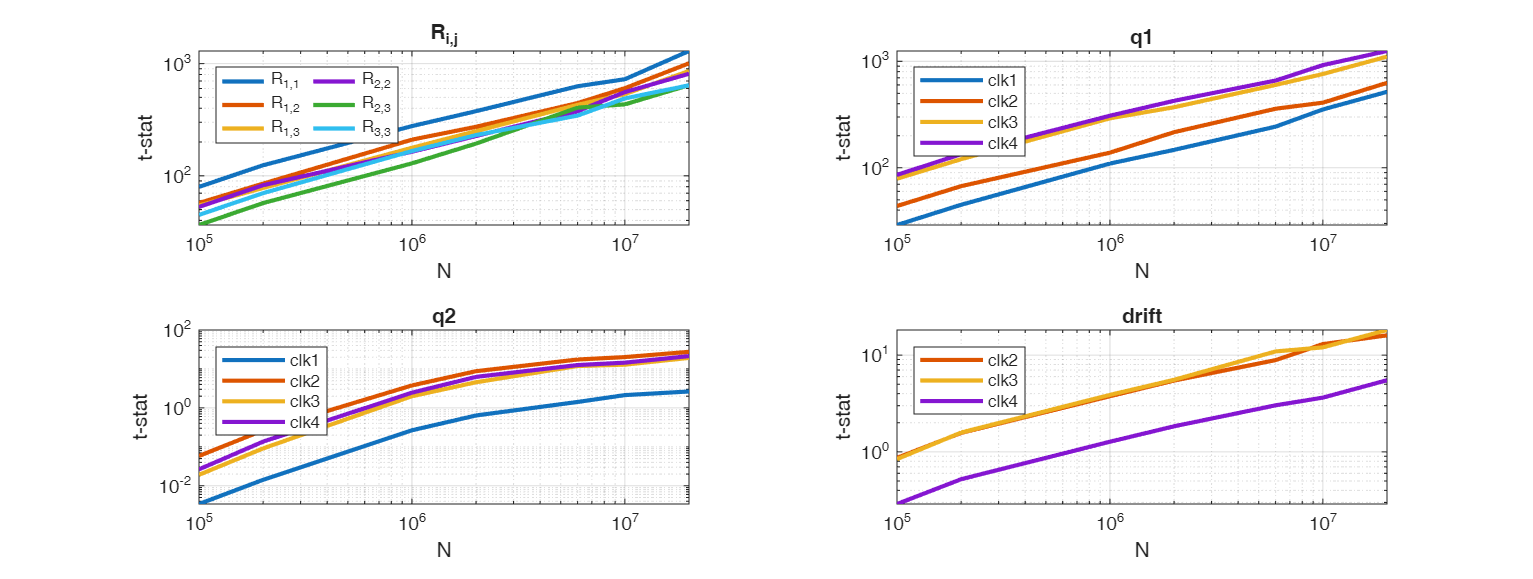}
    \caption{MDM: quality of the noise parameter estimates for different values of $N$  ($\Ts =  \qty{5}{\second}$).}
    \label{fig:tstatDMD}
\end{figure*}

Fig. \ref{fig:tstatHCOV} and \ref{fig:tstatDMD} illustrate the relationship between the quality of individual noise parameter estimates and the number of measurement $N \in \left[ \num{1e5}, \num{1e8} \right]$  (equivalent to the total observation period ranging from 6 days to 3 years). The estimation quality is quantified by a t-statistic, defined as $t_{stat}=\mean (a) / \std( a )$. Both methods yield similar trends for most of the estimated parameters. Statistically significant estimates are obtainable for all parameters $q_1$ and $R_{ij}$ (assuming a confidence level of $0.95$ where $t \ge 2$) even for the minimum duration of $N = \num{1e5}$.
In contrast, for certain $q_2$ (clk1) parameters and drift (clk4), the estimates achieve statistical significance only when $N \ge \num{2e6}$ (equivalent to a measurement period of approximately three months). It is consistent with previous studies \cite{Vernotte2016}, in which the estimation process may produce negative values for $q_2$, a result that is physically non-realizable.

The simulation results validate HCOV and MDM as efficient identification methods offering comparable estimation quality.

The relevant parameters in MDM, as the result of the calculation, are as follows: the resampling periods $\tau_m$ that minimize the covariance of the estimates (Section~\ref{influenceTs} Observation~II) are $\tau_m\!=\!5$\si{\second} for $\bfq_1$ and $\bfr$ (see Fig.~\ref{fig:XiRes2}); $\tau_m\!=\!\num{12180}$\si{\second} for $\bfq_2$ (see Fig.~\ref{fig:XiRes2}); $\tau_m\!=\!\num{9070}$\si{\second} for the estimate of $\bfd$ (see Fig.~\ref{fig:dEst}) and $L\!=\!8$ (see Fig.~\ref{fig:cov_est_vs_L}).

\subsection{Real data}
The historical records of measurements were provided by the ESTEC timing laboratory during year 2022 as part of the ESA project NAVISP-EL1-056. The ensemble consists of triplet AHM atomic clocks iMaser3000 \cite{SAFRAN2022:iMaserDatasheet}. The nominal sampling period of the phase difference measurements is \qty{5}{\second} and the dataset contains $N = \num{1.59e6}$ sampling instants (i.e., this corresponds to a three-months data period). 
The parameter for HCOV is $\te = 20$, where averaging times $\bar{\tau}_{p}$ are evenly spaced in a log-space ranging \qty{5}{\second} to  \qty{7.95E6}{\second}
and 
the parameters for MDM are $S\!=\!100$, $\te\!=\!400$, where resampling period $\bar{\tau}_{p}$ are evenly spaced in a log-space ranging from \qty{5}{\second} to $5\!\cdot\!\!\big\lfloor \tfrac{\num{1589695}}{101\cdot L-2}\big\rfloor $\si{\second}, and $\bar{\bfL}\!=\!\left[ 4, 5, 6, 7, 8, 9, 10, 11\right]$ i.e., $\ell\!=\!8$. 

\begin{figure}[!ht]
    \centering
    \includegraphics[width=0.48\textwidth]{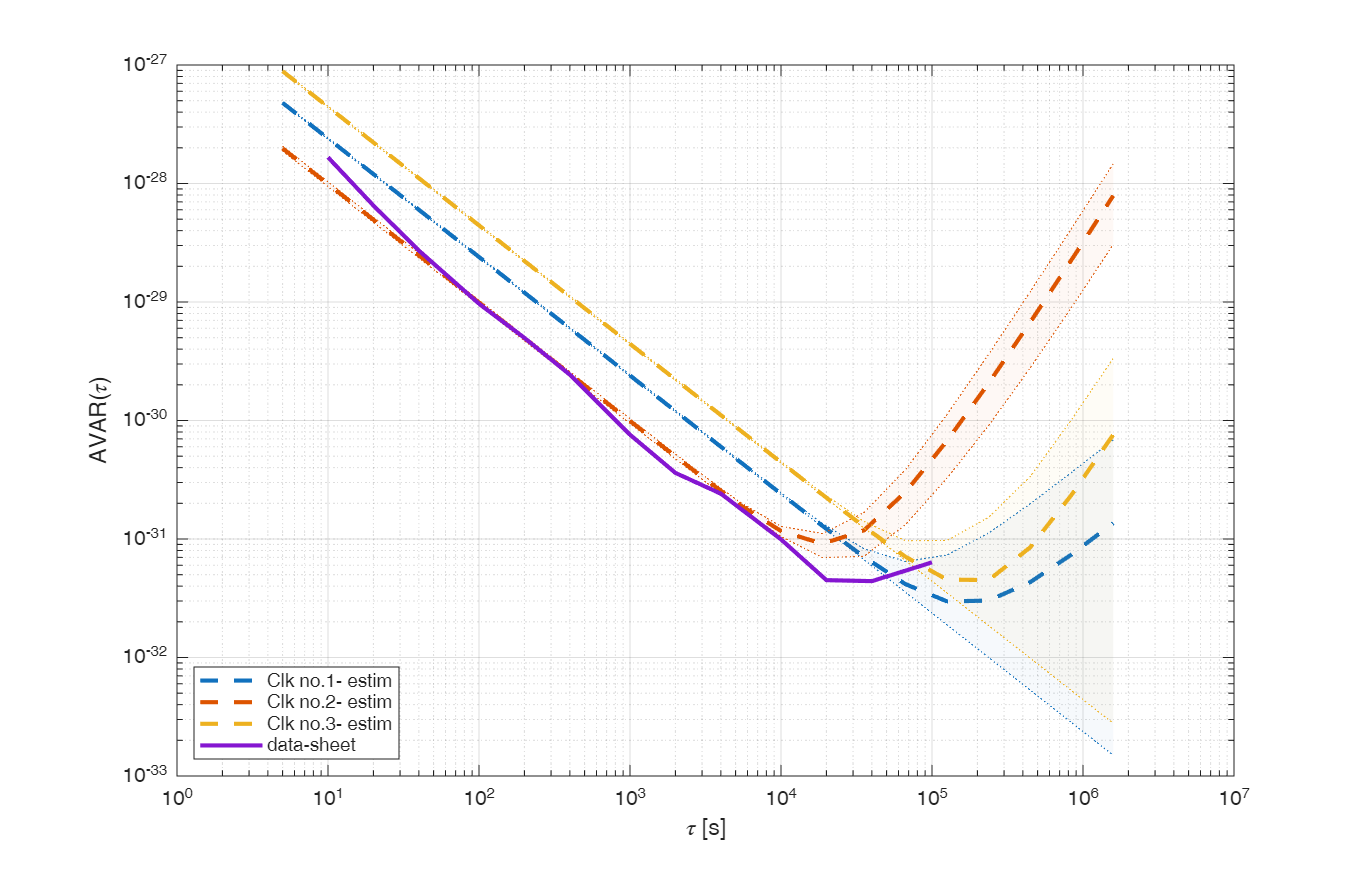}
    \vspace{-0.5cm}
    \caption{AVAR of the iMaser3000 atomic clock adopted from a datasheet and AVAR of estimated clocks (HCOV method - real data).}
    \label{fig:clkACOVreal}
\end{figure}

\begin{figure}[!ht]
    \centering
    \includegraphics[width=0.48\textwidth]{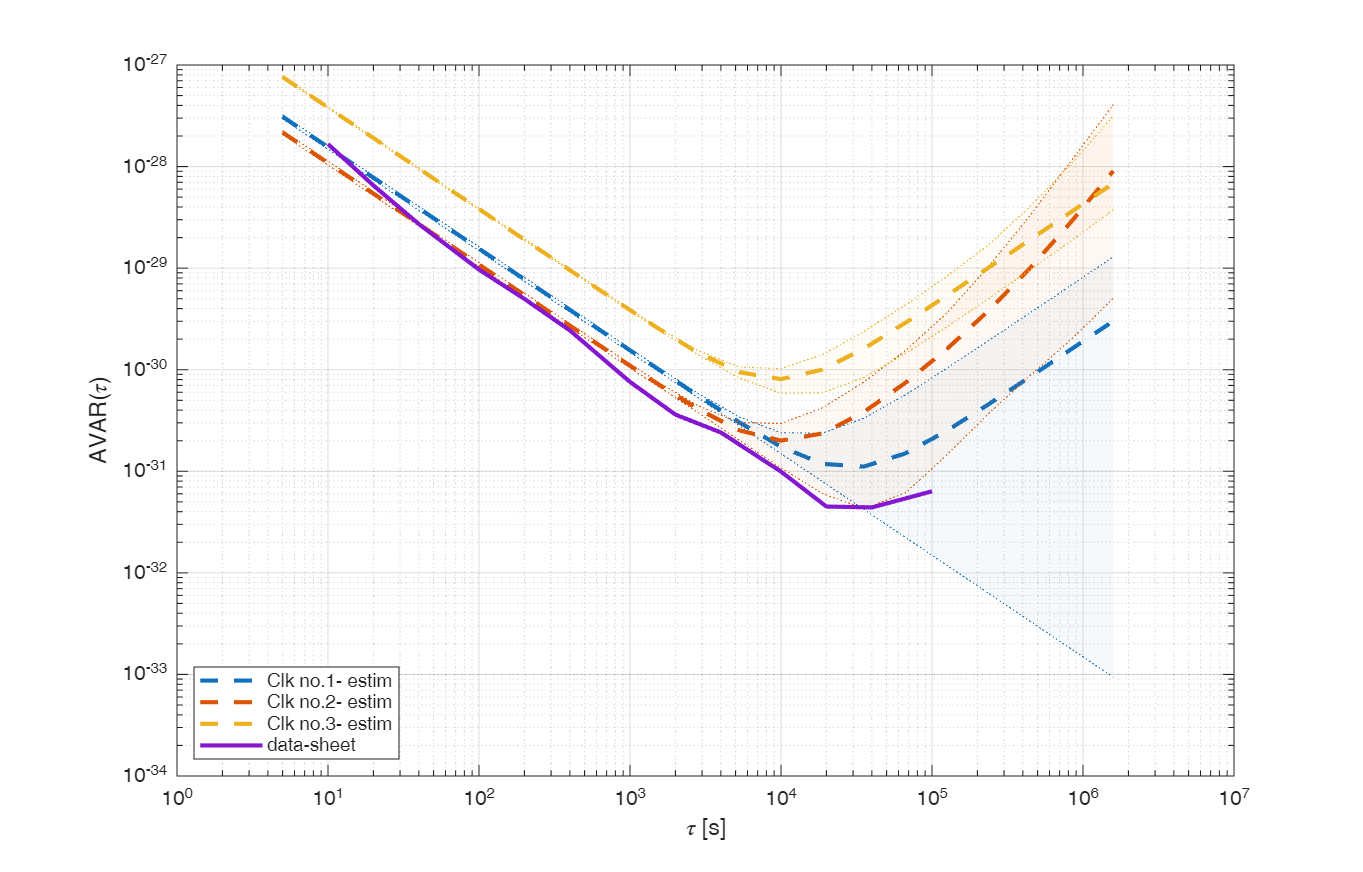}
    \vspace{-0.5cm}
    \caption{AVAR of the iMaser3000 atomic clock adopted from a datasheet and AVAR of estimated clocks (MDM - real data).}
    \label{fig:clkMDMreal}
\end{figure}

The obvious measurement outliers were removed from the real measurements before the parameter estimation methods were applied. The AVARs of individual clocks computed using estimated parameters (dashed lines) and AVAR obtained from data-sheet (solid line) are depicted in Fig.~\ref{fig:clkACOVreal} and \ref{fig:clkMDMreal}. Contrary to the simulation scenario, the parameter estimates obtained by the proposed methods show significant discrepancies. For clocks 1 and 2, the parameter estimates differ by up to \qty{20}{\percent}, whereas for clock 3 the differences span several orders of magnitude. These discrepancies may be attributed to (i) an insufficient amount of data (the available three-month record may be too short, leading to high estimation uncertainty) (ii) the presence of additional noise types in real measurements that are included in the model, and (iii) anomalies in measurements that were not fully removed during preprocessing. However, the results provide an indication of the achievable estimation accuracy under realistic conditions. This insight is particularly relevant for subsequent use of the ensemble model in combination and detection algorithms, which should be designed to remain robust in the presence of such uncertainties.

\begin{table*}[!ht]
    \caption{An overview of real ensemble clocks parameters estimates.}
    \label{Tab:real}
    \centering
    \vspace{-2mm}
    \begin{tblr}{
        colspec={c|c|ccc|ccc|cc|ccc},
        cell{1}{3,6,11} = {c=3}{c},
        cell{1}{9} = {c=2}{c}
        }
         & & $q_1^{(i)}$ $(\times \num{1e-27})$ & & & $q_2^{(i)}$ $(\times \num{1e-36})$ & & & $d^{(i)}$ $(\times \num{1e-21})$ & & $R_{ij}$ $(\times \num{1e-27})$ & & \\ 
         & $i$ or $ij$  & 1 & 2 & 3 & 1 & 2 & 3 &  2 & 3 & 11 & 12 & 22\\
        \hline \hline         
       \SetCell[r=2]{c}  \rotatebox{90}{HCOV}   & $\htheta$ & 2.398 & 0.990 & 4.435 & 0.254 & 4.420 & 0.140 &  7.756 &  0.727 & 8.227 & 6.151 & 7.729  \\[-1.65mm]      
         & Method provided  $\std\! \big(\htheta \big)$ 
         & 0.009 & 0.020 & 0.0170 & 0.233 & 0.820 & 0.255 &  0.769 &  0.227 & 0.021 & 0.022 & 0.035  \\[-1mm]         
         \hline
         \SetCell[r=2]{c}  \rotatebox{90}{MDM} &  $\htheta$  
         & 1.554 & 1.085 & 3.817 & 5.701 & 26.654 & 127.564 &  7.792 & 0.844 & 9.374 & 7.492 & 10.133 &  \\[-1.65mm]    
         &  Method provided  $\std\! \big(\htheta \big)$ 
         & 0.019 & 0.013 & 0.017 & 5.230 & 8.110 & 17.867 & 2.310 & 3.626 & 0.055 & 0.044 & 0.058 &         
    \end{tblr}
\end{table*}





The appropriate parameters in MDM are as follows: the resampling periods $\tau_m$ that minimize the covariance of the estimates (Section~\ref{influenceTs} Observation~II) are: $\tau_m\!=\!5$\si{\second} for $\bfq_1$ and $\bfr$ (see Fig.~\ref{fig:XiRes2}); $\tau_m\!=\!\num{13680}$\si{\second} for $\bfq_2$ (see Fig.~\ref{fig:XiRes2}); $\tau_m\!=\!\num{5600}$\si{\second} for the estimate of $\bfd$ (see Fig.~\ref{fig:dEst}) and $L\!=\!5$.

\section{Discussion} \label{sec:Discussion}
Although the motivations and underlying principles of the proposed methods differ, they share the following features:
\begin{itemize}[topsep=0pt, itemsep=0pt, leftmargin=*]
\item \textit{Independence on state, linear filtering}: Both methods prevent the state dynamics from influencing noise estimates. HVAR can be interpreted alternatively as the output of a specific band-pass filter (rectangular averaging followed by the second-order differentiation) that eliminates the effect of the state (see Appendix~\ref{sec:appendixA}). In MDM, this is achieved through the transformation of the augmented measurement vector \eqref{mpeMDM}.
\item \textit{Noise components decomposition}: Fig.~\ref{fig:XiRes2}, in the case of the MDM, and the HVAR plot, in the case of the HCOV method, can be seen as visualization of different noise types as a function of the resampling period $\tau_m$. However, while the HCOV method  uses regression to fit the HCOV curve at selected points $\bar{\bftau}$ using a model with unknown noise parameters, the MDM first systematically separates the individual noise components and then estimates the noise parameters for the appropriate determined resampling periods $\tau_m$.
\item \textit{Transformation into a linear estimation problem(s)}: HCOV uses the analytical form of HCOV \eqref{eq:acovAnalytic1}, \eqref{eq:acovAnalytic2}, where unknown noise parameters appear linearly. 
In MDM, this is achieved by using a combination of a specific residue 
\eqref{mpeMDM}, its correlation \eqref{noncentMom}, and the replication matrices $\Upsilon_{\bfd}$ \eqref{psid} and $\Upsilon_{\bfq_1}$, $\Upsilon_{\bfq_2}$, $\Upsilon_{\bfr}$ \eqref{Npsi}.

\item \textit{Resampling}: Whereas HCOV and other Allan like statistics can be interpreted as a multiscale resampling variance, MDM uses explicit resampling for both measurements and the model \eqref{eq:ensemble_xx}. 
\item \textit{Bootstrap technique}:  Both methods use bootstrap techniques as statistical resampling methods that estimate the statistical properties of by repeatedly generating new datasets through observed data sampling. Note that the samples are selected deterministically in both cases.
\end{itemize}


\section{Conclusions} \label{sec:Conclusions}
The article presents two identification methods for determining ensemble clocks noise parameters. The performance was evaluated on both synthetic and real data. While the accuracy of the parameter estimation is primarily constrained by sample size, both methods prove to be computationally efficient, ensuring reliable characterization of of noise characteristics under realistic conditions. The proposed methods yield results of similar quality when applied to synthetic data. In a real data case, the results differ, which is due to their different underlying concepts. The choice of method ultimately depends on whether the user prefers a design model fitted to the HCOV plot or a model based on the systematic decomposition of noise into its individual components.

\appendices 
\section{HCOV (HVAR) derivation} \label{sec:appendixA}
HCOV in \eqref{eq:acov_meas} can be rewritten using the state-space representations \eqref{eq:clk1}--\eqref{eq:ensemble_meas} and with the variables rearrangement as
\begin{multline} \label{eq:HcovDer2}
    \sigma^2_{i,j} (\tau_m) = \frac{1}{6\tau_m^2} \mean \Bigg\{ \\
    \bigg[ \left( x_{1,k+3m}^{(i+1)} - 3x_{1,k+2m}^{(i+1)} + 3x_{1,k+m}^{(i+1)} -  x_{1,k}^{(i+1)} \right)  \\ 
     - \left( x_{1,k+3m}^{(1)} - 3x_{1,k+2m}^{(1)} + 3x_{1,k+m}^{(1)} -  x_{1,k}^{(1)} \right)  \\
     + \left( v_{i,k+3m} - 3v_{i,k+2m} + 3v_{i,k+m} -  v_{i,k} \right) \bigg] \\
     \times  \bigg[ \left(x_{1,k+3m}^{(j+1)} - 3x_{1,k+2m}^{(j+1)} + 3x_{1,k+m}^{(j+1)} -  x_{1,k}^{(j+1)} \right) \\ 
     - \left(x_{1,k+3m}^{(1)} - 3x_{1,k+2m}^{(1)} + 3x_{1,k+m}^{(1)} -  x_{1,k}^{(1)} \right)   \\
     + \left(v_{j,k+3m} - 3v_{j,k+2m} + 3v_{j,k+m} -  v_{j,k} \right) \bigg] \Bigg\}.
\end{multline}
This can be rewritten using the third difference of variable $g_k$ denoted as $\Delta^3 g_{k} = g_{k+3m} - 3g_{k+2m} + 3g_{k+m} - g_{k}$ as
\begin{multline}  \label{eq:HcovDer3}
     \sigma^2_{i,j} (\tau_m) = \frac{1}{6\tau_m^2} \mean \left\{  \left( \Delta^3 x_{1,k}^{(i+1)} -  \Delta^3 x_{1,k}^{(1)}  + \Delta^3 v_{i,k} \right) \right. \\
     \times \left. \left( \Delta^3 x_{1,k}^{(j+1)} -  \Delta^3 x_{1,k}^{(1)}  + \Delta^3 v_{j,k} \right) \right\}.
\end{multline}  
Considering that the individual ensemble clocks are uncorrelated (as follows from the block diagonal form of the CM $\bfQ$) and using $\mean(x\cdot y) = \text{cov}(x,y)+\mean(x)\mean(y)$, the equation \eqref{eq:HcovDer3} can be simplified to the form
\begin{multline}   \label{eq:HcovDer4}
    \sigma^2_{i,j} (\tau_m) =  \frac{1}{6\tau_m^2} \left(  \var \left(\Delta^3 x_{1,k}^{(1)} \right)  + \cov  \left(  \Delta^3 v_{i,k} \Delta^3 v_{j,k} \right)  \right.\\
     + \mean \left(  \Delta^3 x_{1,k}^{(i+1)} - \Delta^3 x_{1,k}^{(1)}  + \Delta^3 v_{i,k}  \right) \\
      \left. \times \mean \left(  \Delta^3 x_{1,k}^{(j+1)} -  \Delta^3 x_{1,k}^{(1)}  + \Delta^3 v_{j,k} \right)\right).
\end{multline}
Before focusing on individual members in equation~\eqref{eq:HcovDer4}, it is convenient to introduce the relations
\begin{align}
    \bfx_{k+m}^{(i)} &= \bfF^m \bfx_k^{(i)} + \bfw_m^{(i)}, \label{eq:ssM} \\
    \quad \bfx_{k+2m}^{(i)} &= \bfF^m \bfx_{k+m}^{(i)} + \tilde{\bfw}_m^{(i)}, \label{eq:ssM2} \\ 
    \quad \bfx_{k+3m}^{(i)} &= \bfF^m \bfx_{k+2m}^{(i)} + \doubletilde{\bfw}_m^{(i)}, \label{eq:ssM3}
\end{align}
where $\bfF^m = 
    \begin{bmatrix} 
      1 & m \Ts \\ 0 & 1 
    \end{bmatrix}$, 
white Gaussian noises $\bfw_m^{(i)}$, $\tilde{\bfw}_m^{(i)}$ and $\doubletilde{\bfw}_m^{(i)}$  are independent and have identical statistical properties, i.e., $\bfw_m^{(i)}, \tilde{\bfw}_m^{(i)},  \doubletilde{\bfw}_m^{(i)}  \sim \mathcal{N}\left(\bfmu_m^{(i)}, \bfQ_m^{(i)}\right)$
with 
\begin{align}
    \bfmu_m^{(i)} &= d^{(i)}
    \!\begin{bmatrix}  
       m^2 \Ts^2/2 \\
       m \Ts
     \end{bmatrix}, \\ 
    \bfQ_m^{(i)} &= 
    \!\begin{bmatrix} 
      q_1^{(i)} m \Ts + q_2^{(i)} \Ts^3 m^3/3 & q_2^{(i)} m^2 \Ts^2/2 \\
      q_2^{(i)} m^2 \Ts^2/2 & q_2^{(i)} m \Ts 
    \end{bmatrix}. 
\end{align}
Note that \eqref{eq:ssM}--\eqref{eq:ssM3} follows from the fact that $\bfx_{k+m}$ can be understood as a change in the sampling period of the dynamic system \eqref{eq:ensemble_state}--\eqref{eq:ensemble_meas} by a multiple of $m$. 

Now, the individual members of expression \eqref{eq:HcovDer4}  can be covered. The first one is concerned with $\Delta^3 x_{1,k}^{(i)}$ that can be rewritten using equations~\eqref{eq:ssM}--\eqref{eq:ssM3} as follows 
\begin{align}
\setlength{\arraycolsep}{1.5pt}
    \begin{aligned}
    \!\!\!\!\Delta^3 x_{1,k}^{(i)} &= x_{1,k+3m}^{(i)} - 3x_{1,k+2m}^{(i)} + 3x_{1,k+m}^{(i)} - x_{1,k}^{(i)} \\
    & = \begin{bmatrix}  1 & 0 \end{bmatrix} \left(\bfx_{k+3m}^{(i)} - 3\bfx_{k+2m}^{(i)} + 3\bfx_{k+m}^{(i)} - \bfx_k^{(i)} \right)  \\
    & = \begin{bmatrix}  1 & 0 \end{bmatrix} \left( \left(\bfF^{3m} - 3\bfF^{2m} + 3\bfF^m - \bfI \right) \bfx_k^{(i)}  \right.\\
     &\hspace{-6mm} \!+\! \left. \left( \bfF^{2m} - 3\bfF^{m} + 3\bfI \right) \bfw_m^{(i)} \!+\! \left( \bfF^{m}-3\bfI \right) \tilde{\bfw}_m^{(i)} + \doubletilde{\bfw}_m^{(i)} \right)\!.\hspace{-3mm}
    \end{aligned}
\end{align}
It is clear that $(\bfF^{3m} - 3\bfF^{2m} + 3\bfF^m - \bfI)$ is a zero matrix. Thus, $\Delta^3 x_{1,k}^{(i)}$ depends solely on the noise characteristics as
\begin{align}
    \!\!\!\!\Delta^3 x_{1,k}^{(i)} \!&=\!  \begin{bmatrix}  1 & 0 \end{bmatrix} \left( \begin{bmatrix} 1 & -m \Ts \\ 0 & 1 \end{bmatrix} \bfw_m^{(i)} \!+\!  \begin{bmatrix} -2 & m \Ts \\ 0 & -2 \end{bmatrix}\tilde{\bfw}_m^{(i)}  \!+\! \doubletilde{\bfw}_m^{(i)} \right) 
    \nonumber\\
    \!&=\!  \begin{bmatrix}1 & -m \Ts\end{bmatrix} \bfw_m^{(i)} +  \begin{bmatrix}-2 & m \Ts\end{bmatrix} \tilde{\bfw}_m^{(i)}  + \begin{bmatrix}1 & 0\end{bmatrix} \doubletilde{\bfw}_m^{(i)}.\!\!\!\!
\end{align}
One can easily obtain mean and variance of $\Delta^3 x_{1,k}^{(i)}$ as
\begin{align} 
    \!\!\!\!\mean (\Delta^3 x_{1,k}^{(i)}) \!=&\! \left( \begin{bmatrix}1 & -m \Ts\end{bmatrix}  \!+\! \begin{bmatrix}-2 & m \Ts\end{bmatrix} \!+\!  \begin{bmatrix}1 & 0\end{bmatrix} \right)\! \bfmu^{(i)}_m \!=\! 0\!\!\!\! \label{eq:ED3x}
    \\
    \!\!\!\!\var(\Delta^3 x_{1,k}^{(i)}) \!=&\!\begin{bmatrix} 1 & -m \Ts\end{bmatrix} \!\bfQ_m^{(i)}(\cdot)\T
    +\!\begin{bmatrix} -2 &\ m \Ts\end{bmatrix} \!\bfQ_m^{(i)} (\cdot)\T
    \nonumber\\
    &\hspace{10mm}+\! \begin{bmatrix} 1 & 0\end{bmatrix} \!\bfQ_m^{(i)} (\cdot)\T
    \!= 6 q_1^{(i)} \tau_m  \!+\! q_2^{(i)} \tau_m^3.\!\!\label{eq:varD3x}
\end{align}
Next, let's focus on $\Delta^3 v_{i,k}$. Since
\begin{equation}
     \Delta^3 v_{i,k} = v_{i,k+3m} - 3v_{i,k+2m} +3v_{i,k+m} + v_{i,k}
\end{equation}
and $v_k$ is an independent sequence, for mean and variance of $\Delta^2 v_{i,k}$  can be stated
\begin{align}
     \mean \left( \Delta^3 v_{i,k}  \right) &= 0 \label{eq:ED3v}
     , \quad 
     \var (\Delta^3 v_{i,k}) =  
     20R_{ii}.
\end{align}
Finally, the covariance $\cov(\Delta^3 v_{i,k},\Delta^3 v_{j,k})$ can be deduced as
\begin{equation}   \label{eq:covD3v}
     \cov(\Delta^3 v_{i,k},\Delta^3 v_{j,k})=  R_{ij} \!+\! 9R_{ij} \!+\! 9R_{ij} \!+\! R_{ij} \!=\! 20R_{ij}.
\end{equation}
Using \eqref{eq:ED3x}, \eqref{eq:varD3x}, \eqref{eq:ED3v} and \eqref{eq:covD3v}, HCOV $\sigma^2_{i,j} (\tau_m)$ from \eqref{eq:HcovDer4} results in \eqref{eq:acovAnalytic1}.
Finally, HVAR $\sigma^2_{i,i} (\tau_m)$ given by \eqref{eq:acovAnalytic2} can be obtained analogously by the procedure \eqref{eq:acov_meas} and \eqref{eq:HcovDer2}--\eqref{eq:covD3v}.

\section{Sample-Based Covariance Estimation} \label{sec:appendixB}
Consider the following linear equation
\begin{align}
    \bfalpha_k &= \bfD \bfomega + \bfpi_k, k=1,2,\cdots,N,
\end{align}
where $\bfalpha_k$ is a known measurement, $\bfD$ is a known matrix, $\bfomega$ is a vector of unknown parameters, and $\bfpi_k$ is zero mean noise with CM $\cov(\bfpi_{k+r},\bfpi_k)=\bfPi_r\!\neq\!\bfnul$ for $\abs(r)\!<\! L$ and $\bfPi_r\!=\!\bfnul$ for $\abs(r)\!\geq\! L$.

The estimate of $\bfomega$ using LS for a total of $N$ available measurements is the following
\begin{align}
    \!\!\!\widehat{\bfomega_N}
    &=\!\!\left(\!\bfD\sT
    \begin{bsmallmatrix} \!\bfI & \!\cdots\! & \bfI \!\end{bsmallmatrix}\!
    \begin{bsmallmatrix} \bfI \\[-1mm] \vdots \\ \bfI \end{bsmallmatrix}
    \!\bfD\!\right)^{\!\!\!-1}
    \!\!\!\!\bfD\sT
    \begin{bsmallmatrix} \!\bfI& \!\cdots\! & \bfI\! \end{bsmallmatrix}\!
    \begin{bsmallmatrix} \bfalpha_1 \\[-1mm] \vdots \\ \bfalpha_N \end{bsmallmatrix}\!\!
    =
    \!\bfD^{\dagger}\tfrac{1}{N}\textstyle\sum_{k=1}^{N}\bfalpha_k.\!\!\label{estApen}
\end{align}
The CM  of the estimate \eqref{estApen} is determined by the following 
\begin{subequations}
\begin{align}
    \!\!\!\!\cov(\widehat{\bfomega_N})
    &=\!\bfD^{\dagger}\tfrac{1}{N^2}\textstyle\sum_{k=1}^{N}\!\sum_{i=1}^{N}\cov\left(
    \bfpi_k,
    \bfpi_i\right)\left(\bfD^{\dagger}\right)^{\ssT}
    \\
    &=\!\bfD^{\dagger}\!\!\left( 
    \begin{matrix}
        \frac{N\bfPi_0}{N^2}+\frac{(N-1)(\bfPi_1+\bfPi_1\T)}{N^2}+
        \\[1mm]
        +\hdots+\!\frac{(N-(L-1))(\bfPi_{L-1}+\bfPi_{L-1}\T)}{N^2}\!\!
    \end{matrix} \ \!\!\right)\!\!\left(\bfD^{\dagger}\right)^{\ssT}\!\!\!.\!\!\label{covEstApen}
\end{align}
\end{subequations}

It can be shown that the matrix inequality $2\bfPi_0\!\succeq\!(\bfPi_1\!+\!\bfPi_1\T)\!\succeq\!\cdots\!\succeq\!(\bfPi_L\!+\!\bfPi_L\T)\!=\!\bfnul$ holds
i.e., the dependence $\cov(\bfpi_{k+r},\bfpi_k)$ decreases as $\abs(r)$ increases from $0$ to $L$.
Based on this, the user chose a linearly decreasing dependency as the most appropriate compromise between the absence of dependency (i.e., $\!\bfPi_0\!\neq\!\bfnul$ and $\!\bfPi_1\!\!=\!\!\bfPi_2\!\!=\!\!\cdots\!\!=\!\!\bfPi_L\!\!=\!\!\bfnul$) and full dependency (i.e., $2\bfPi_0\!=\!(\bfPi_1\!\!+\!\!\bfPi_1\T)\!=\!\cdots\!=\!\!(\bfPi_{L-1}\!\!+\!\!\bfPi_{L-1}\T)$ and $\bfPi_L\!=\!\bfnul$).
Therefore the estimate $\big(\widehat{\bfPi_r\!\!+\!\bfPi_r\T}\big) \!=\! 2\bfPi_0\frac{L-r}{L}$ for $0\leq \abs(r)\leq L$ is used. Using this estimate in \eqref{covEstApen}, the estimate of the CM  of the estimate can be obtained by
\begin{subequations}
\begin{align}
    \widehat{\cov(\widehat{\bfomega_N})}
    &=\!\bfD^{\dagger}\!\left( 
    \begin{matrix}
        \frac{N\bfPi_0}{N^2}+\frac{2(N-1)\bfPi_0\frac{L-1}{L}}{N^2}+
        \\[1mm]
        +\hdots+\frac{2(N-(L-1))\bfPi_0\frac{1}{L}}{N^2}
    \end{matrix} \ \right)\!\!\left(\bfD^{\dagger}\right)^{\ssT}
    \\
     &=\!\bfD^{\dagger}\bfPi_0 \left(\bfD^{\dagger}\right)^{\ssT}
    \underbrace{\!\left(\! \tfrac{1}{N}\!+\!\textstyle\sum_{i=1}^{L-1}\tfrac{2(N-i)}{N^2}\tfrac{L-i}{L} \right)\!}_{\calC(N,L)}.\label{covEstAprApen}
\end{align}
\end{subequations}
The relation \eqref{covEstAprApen} is useful when the estimate od the CM  of the estimate $\widehat{\cov(\widehat{\bfomega_N})}$ for a total of $N$ measurements is available, and the CM  of the estimate $\cov(\widehat{\bfomega_M})$ for a total of $M$ measurements is needed. The estimate can be obtained by
\begin{align}
    \widehat{\cov(\widehat{\bfomega_M})}
    =\widehat{\cov(\widehat{\bfomega_N})}\tfrac{\calC(M,L)}{\calC(N,L)}.
\end{align}

\bibliographystyle{IEEEtran}
\bibliography{literatura_ivo}

\end{document}